\documentclass[final,5p,twocolum]{elsarticle}

\usepackage{longtable}
\usepackage{graphics}
\usepackage{graphicx}
\usepackage{epsfig}
\usepackage{amssymb}
\usepackage{amsthm}
\usepackage{lineno}
\usepackage{setspace}
\usepackage{lscape}
\usepackage{xcolor}
\usepackage{lscape}

\journal{?}

\begin{document}

\begin{frontmatter}

\author{Jafar Ghazanfarian\corref{cor1}}
\ead{j.ghazanfarian@znu.ac.ir}
\address{Mechanical Engineering Department, Faculty of Engineering, University of Zanjan, P.O. Box 45195-313, Zanjan, Iran.}
\cortext[cor1]{Corresponding author, Tel.: +98(241) 3305 4142. All authors contributed equally to this work.}
\author{Mohammad Mostafa Mohammadi}
\address{Mechanical Engineering Department, Faculty of Engineering, University of Zanjan, P.O. Box 45195-313, Zanjan, Iran.}
\author{Kenji Uchino}
\address{International Center for Actuators and Transducers, The Pennsylvania State University, University Park, PA, 16802, USA}

\title{Piezoelectric Energy Harvesting: a Systematic Review of Reviews}

\begin{abstract}
In the last decade, an explosive attention has been paid to piezoelectric harvesters due to their flexibility in design and increasing need to small-scale energy generation. As a result, various energy review papers have been presented by many researchers to cover different aspects of piezoelectric-based energy harvesting, including piezo-materials, modeling approaches, and design points for various applications. Most of such papers tried to shed light on recent progresses in related interdisciplinary fields, and to pave the road for future prospects of development of such technologies. However, there are some missing parts, overlaps, or even some contradictions in the review papers. In the present review of review articles, recommendations for future research directions suggested by the review papers have been systematically summed up under one umbrella. At the final section, topics for missing review papers, concluding remarks on outlooks and possible research topics, and strategy-misleading contents have been presented. The review papers have been evaluated based on merits and subcategories and authors' choice papers have been presented for each section based on clear classification criteria.
\end{abstract}

\end{frontmatter}

\section*{Highlights}
\begin{itemize}
   \item A comparative overview of reviews in the map of piezo-energy harvesting is presented.
   \item An extensive description of research lines for future research is provided.
   \item Classification of reviews is presented based on subcategories and merits.
   \item  Authors' choice papers are presented for each section.
\end{itemize}

\section*{Keywords}
Energy harvesting; piezoelectric; energy conversion; renewable energies; micro-electro-mechanical systems

\tableofcontents

\section{Introduction}
Due to recent developments of portable and wearable electronics, wireless electronic systems, implantable medical devices, energy-autonomous systems, monitoring systems, and MEMS/NEMS-based devices, the procedure of small-scale generation of energy may lead to a revolution in development of compact power technologies.

\begin{figure}[htbp] \setlength{\unitlength}{1mm}
\hspace{-0.0in}
\begin{picture}(90,55)
\includegraphics[height=55mm]{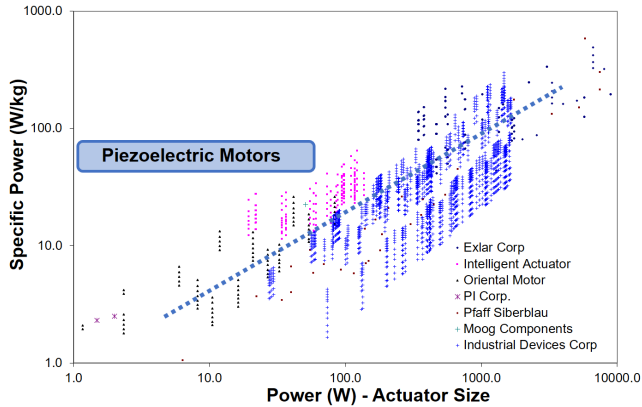}
\end{picture}
\caption {Comparison of the specific power with respect to the power~\cite{uchino}.}
\label{Fig:com}
\end{figure}

Figure~\ref{Fig:com} presents the output power density variation versus the actual motor power for 2000 commercial electromagnetic motors. Electromagnetic motors are superior for the production of power levels higher than 100 W. However, because the efficiency is significantly dropped below 100 W, the piezoelectric devices with power density insensitive to their size will replace battery-operated small portable electronic equipment less than 50 W level. It is not logical to compare the energy harvesting systems with the MW power level. Hence, it is necessary for researchers to determine their original piezo-harvesting target, which should be basically the replacement of compact batteries, one of the toxic wastes in the sustainable society~\cite{uchino}.

Dutoit et al.~\cite{Dut2005} provided a comparison based on the density of the output power, and indicated that the power densities of the fixed-energy density sources extensively drop after just 1 year of operation. So, they need maintenance and repair if possible. Designing an effective power normalization scheme, strain cancelation due to multiple input vibration components, optimizing the minimum vibration level required for positive energy harvesting, and the prototype testing to eliminate the proof mass are among the suggestions as future works.


Advantages of the piezoelectric energy harnessing include simple structure without several additional components, no need to moving parts or mechanical constraints, environment friendliness and being ecologically safe, portability, coupled operation with other renewable energies, no need to an external voltage source, compatibility with MEMS, easy fabrication with microelectronic devices, reasonable output power density, cost effectiveness, and scalability. Hence, piezo-materials are an excellent candidate to replace batteries with short lifespan for powering macro to nanoscale electronic devices.

Piezo-materials can extract power directly from structural vibrations or other environmental mechanical waste energy sources in infrastructures (bridges, buildings), biomedical systems, health care and medicine, and they can be used for transducers, actuators, and surface acoustic wave device operation. Some disadvantages of the piezo-harvesters are high output impedance, producing relatively high output voltages at low electrical current, and rather large mechanical impedance.

The number of review papers on piezoelectric energy harvesting has been extensively increased in the recent decade. Due to the tremendous number of published review papers in this field, finding an appropriate review paper became challenging. On the other hand, there are lots of overlaps, similarities, missing parts, and sometimes contradictions between different reviews. Therefore, the main motivation of the present paper is to present a systematic review of the review papers on piezoelectric energy harvesting. We tried to summarize all deficits, advantages, and missing parts of the existing review papers on piezo-energy harvesting systems.

An extensive search among database sources identified 91 review papers in diverse applications related to the piezoelectric energy harvesting. As will be demonstrated later, such papers have  present different concluding remarks for the area of usage, materials, design approaches, and mathematical models. We tried to perform a very detailed searching procedure with several keywords and search engines to cover all published review papers, and to find the review papers without "piezo" directly mentioned in the title.

\begin{figure}[htbp] \setlength{\unitlength}{1mm}
\begin{center}
\begin{picture}(70,75)
\includegraphics[height=75mm]{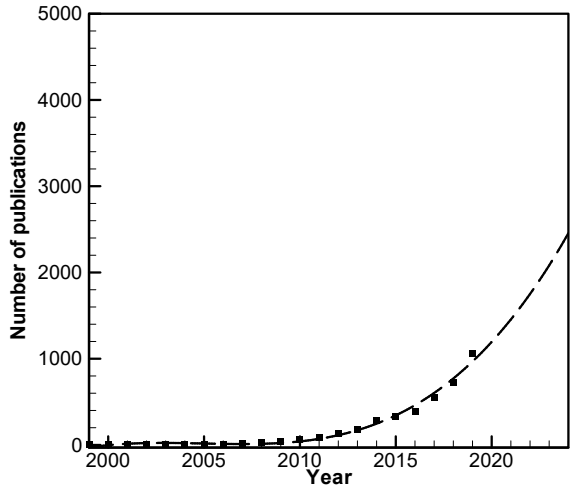}
\end{picture}
\end{center}
\caption {Statistics and future estimation of publications on piezoelectric energy harvesting.}
\label{Fig1:stat}
\end{figure}

The statistics of publications during two recent decades excluding conference papers, extracted using the keyword "piezo AND energy harvesting" from SCOPUS are shown in Fig.~\ref{Fig1:stat}. The results from SCOPUS included the overall number of 4435 documents, containing 874 open access papers, 130 book chapters, and 36 books. The national natural science foundation of China, the fundamental research funds for the central universities, and the national research foundation of Korea were the most frequent funding sponsors. Most common subject areas were engineering, material sciences, physics and astronomy, chemistry, and energy. An extrapolation shown in the figure anticipates publication of about 2500 articles per year during the coming three years.

\begin{figure*}[htbp] \setlength{\unitlength}{1mm}
\begin{center}
\begin{picture}(70,250)
\hspace{-4cm}
\includegraphics[height=250mm]{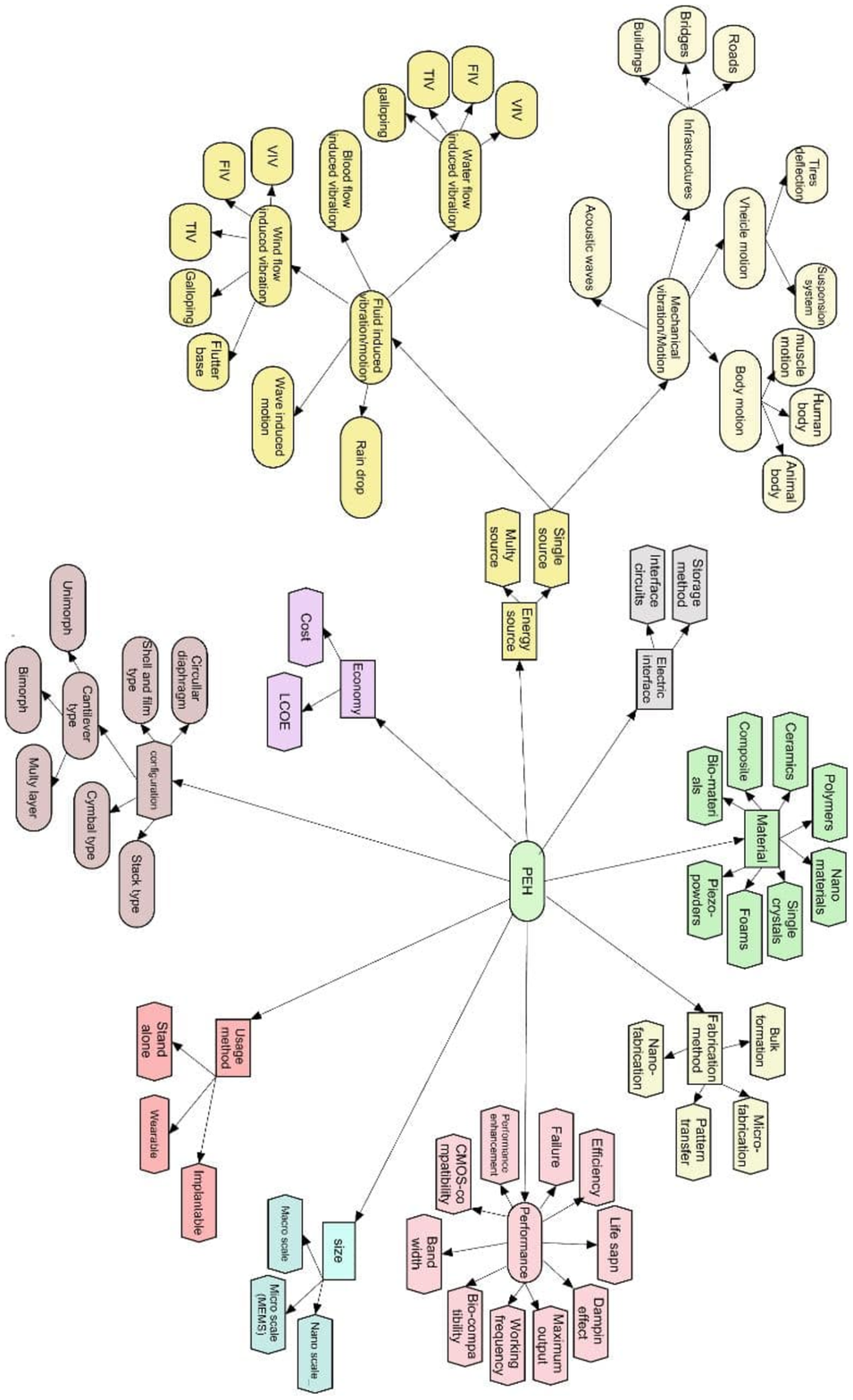}
\end{picture}
\end{center}
\caption {Strategic map of piezoelectric energy harvesting design aspects, modeling approaches, and applications.}
\label{Fig:map}
\end{figure*}

Due to interdisciplinary nature of piezoelectric energy harvesting, prediction of behavior of the piezo-generators are related to different thermo-electro-mechanical sciences as well as material engineering. We have illustrated a systematic map of various aspects of piezo-energy harvesting  in Fig.~\ref{Fig:map}. Different branches of connected sciences and applications include fabrication methods, hybrid systems, performance evaluation, size, utilization methods, configurations, modeling aspects, economical points, energy sources, optimization, design of an electric interface, and selection of proper materials. All sub-branches in the figure will be discussed in subsections of the present paper.


A review article is not an omnibus of the paper collection. The review should be written for criticizing or praising each paper. Evaluation of the review papers and their contribution to the field will be presented based on the following criteria:
\begin{enumerate}
\item Having a solid evaluation philosophy by the reviewer.
\item Presenting non-general future research directions in the summary/conclusion of the paper.
\item Paying attention to the critical design aspects such as electromechanical coupling factor or actual resonance frequency.
\item Many papers report the harvesting energy around the resonance range. Though the typical noise vibration is in a much lower frequency range, the researchers measure the amplified resonance response (even at a frequency higher than 1 kHz).
\item If the harvested energy is lower than 1mW, which is lower than the required electric energy to operate a typical energy harvesting electric circuit with a DC/DC converter (typically around 2-3mW), it is somehow difficult to describe the system as an energy harvesting device.
\item The complete energy flow or exact efficiency from the input mechanical noise energy to the final electrical energy in a rechargeable battery via the piezoelectric transducer is an important part of the review from the applicational/industrial viewpoint.
\item Number of sub-fields covered in the review paper.
\item The review papers may provide enough theoretical background of piezoelectric energy harvesting, practical material selection, device design optimization, energy harvesting electric circuits to help readers prevent the "Google syndrome"~\cite{kenji2021}.
\end{enumerate}

The scoring strategy is as follows: 1 point for the number of conclusions reported, 1 point for the number of sub-categories covered, 2 points for paying attention to the merits, and 1 point for reporting the minimum required energy output level. Details of scores for each parts is presented in the tables inside brackets. The reviews with the scores between 0-1, 1-2, 2-3, 3-4, 4-5, respectively are labeled with E to A. It should be noted that the value of minimum required output should be clearly addressed among concluding remarks, conclusions, future directions, abstract, or introduction.

The outline of the paper is as follows. At the first section, the focus is on the reviews about the design process, structure, material considerations, size effects, and the mathematical modeling challenges. At the second part of the article, the main theme will be evaluating applications of the piezo-harvesters. The most common applications include vibrational energy sources, fluid-based harvesters, scavenging energy from ambient waste energies, and energy harnessing in biological applications. In the last section, a summary of future challenges, research directions, and missing review topics will be presented.

\section{Reviews with non-focused topics}
The discussed papers in this section are general review articles without having a specific focal point. Safaei et al.~\cite{Safaei2019} presented a review of energy harvesting using piezoelectric materials during 2008 to 2018. This article is an update of their previous review~\cite{anton2007}, and covers lead-free piezo-materials, piezoelectric single crystals, high-temperature piezoelectricity, piezoelectric nanocomposites, piezoelectric foams, nonlinear and broadband transducers, and micro-electro-mechanical transducers. They also discussed several types of piezoelectric transducers, the mathematical modeling, energy conditioning circuitry, and applications such as fluids, windmill-style harvesters, flutter-style harvesters, from human body, wearable devices, implantable devices, animal-based systems, infrastructure, vehicles, and multifunctional/multi-source energy harvesting. Several useful illustrations have been presented in the paper, which sum up different technologies in a unified framework. However, their brief recommendations for future horizons in the field, including fabrication of piezoelectric nanofibers, piezoelectric thin films, printable piezoelectric materials, exploiting internal resonance of structures, and development of metamaterials and metastructures may be extended to cover other aspects presented in Tab.~\ref{general11}.

Anton and Sodano~\cite{anton2007} reviewed some general topics published between 2003 to 2006, discussing the efficiency improvement, configurations, circuitry and method of power storage, implantable and wearable power supplies, harvesting from the ambient fluid flows, the micro-electro-mechanical systems, and the self-powered sensors without a clear classification. They described that the future directions are the development of a complete self-powered device that includes a combination of power harvester, storage, and application circuitry. Also, they declared that the enhancement of energy generation and storage methods along with 	decreasing the power requirements of electronic devices may be a prime target.
Taware and Deshmukh~\cite{Taw2013} briefly reviewed a number of literature in the field of piezoelectric energy harvesting. They mentioned advantages and disadvantages of some of piezoelectric materials. They explained the cantilever-based piezoelectric energy harvesters, their related design points, and mathematical modeling.
Khaligh et al.~\cite{Khaligh2009} addressed the piezoelectric and electromagnetic generators suitable for human-powered and vibration-based devices, including resonant, rotational, and hybrid devices.
A brief information has been presented about the hybrid generators provided by an imbalanced rotor that needs more in-deep investigations in future reviews. 
Sharma and Baredar~\cite{sharma2019} analyzed the current methods to harvest energy from vibration using a piezoelectric setup in low-range frequency zone by analyzing piezoelectric material properties based on modeling and experimental investigations.
They indicate that the disadvantages of the piezo-harvesters are depolarization, sudden breaking of piezo layer due to high brittleness and poor coupling coefficient, poor adhesive properties of PVDF material, and lower electromagnetic coupling coefficient of PZT. They discussed that the design of high-efficiency energy harvesters, invention of new energy harvesting designs by exploring non-linear benefits, and design of portable compact-size systems with integrated functions are forthcoming challenges.

Mateu and Moll~\cite{Mateu2005} presented an overview of several methods to design an energy harvesting device for microelectronics depending on the type of available energy. They summarized the power consumption of the microelectronic devices and explained the working principals of piezoelectric, electrostatic, magnetic induction, and electromagnetic radiation-based generators. 
Calio et al.~\cite{calio2014} reviewed the material properties of about 19 piezo-materials, the piezo-harvesters operating modes, resonant/non-resonant operations, optimal shape of the beam, the frequency tuning, the rotational device configurations, the power density and bandwidth, and the conditioning circuitry. They tried to present a selection guide between piezoelectric materials based on the power output and the operating modes. They concluded that the resonant $d_{33}$ cantilever beam needs to be optimized and $d_{15}$ harvester is still too complex to be fabricated but has great potentials. This paper may be a good suggestion for beginners to start a research in the field of piezoelectric energy harvesting.
Batra et al.~\cite{Batra2016} reviewed mathematical modeling and constitutive equations for piezo-materials, the lumped parameter modeling, mechanisms of piezoelectric energy conversion, and operating principles of the piezoelectric energy harvesters. 
Sun et al.~\cite{Sun2012} made a review on applications of piezoelectric harvesters. However, they put everything in a nutshell. Such topics need more close considerations. 
A super short review paper exists~\cite{Sharm2018} that mainly has focused on some points about the history of piezoelectric effect, piezo-materials, and applications like harvesting from footsteps and roads.


\begin{table*}
\caption{{\normalsize Overall evaluation of review papers written about non-focused topics on piezoelectric energy harvesting. "Cons." stands for conclusions. Numbers in brackets are scores for each item.\\
\textbf{Conclusions:} 1 Efficiency/performance improvement, 2 frequency tuning, 3 safety issues, 4 costs, hybrid harvesters, 5 non-linear models, 6 battery replacement, 7 miniaturization, 8 steady operation, 9 more efficient materials. \\
\textbf{Merits:} 1: electromechanical coupling factor, 2: realistic resonance, 3: energy flow, 4: paying attention to the range of output power. \\
\textbf{Sub-categories:} 1: microscale, 2: electrostatic, 3: magnetic induction, 4: electromagnetic radiation, 5: thermal energy, 6: circuit, 7: wearable device, 8: ambient fluid flow, 9: sensors, 10: material, 11: human, 12: vibration, 13: hybrid device, 14: modelling, 15: material, 16: road and shoe, 17: fluids, 18: animals.}
}
\resizebox{1.0 \textwidth}{!}{
\begin{tabular}{c|p{2.5cm}|p{1.2cm}|p{1.5cm}|p{2cm}|p{2.5cm}|p{1.0cm}|p{4.0cm}} \hline\hline
\# Cons. & Minimum required output & \# Refs. & Merits & Sub-categories & Ref. & Grade & Highlights \\ \hline\hline
6 (0.67)& ~$\mu W$ to $mW$ (1) & 478 & 1, 2, 3, 4 (2.00) & 1, 6, 11, 14, 15, 16, 17, 18 (0.44) & Safaei et al.~\cite{Safaei2019}& A &High-temperature devices, metamaterials \\\hline 
5 (0.56)& ~$\mu W$ (1) & 90 &1, 3, 4 (1.5)&1, 6, 7, 8, 9, 10 (0.33)&Anton and Sodani~\cite{anton2007}& B &-\\\hline 
5 (0.56)&  375$\mu W$ (1)&14&2, 4 (1.0)&11, 12, 14 (0.17)&Taware and Deshmukh~\cite{Taw2013}& C & -\\\hline 
3 (0.33)&  ~$\mu W$ to $mW$ (1) &54&1, 4 (1.0)&2, 11, 12, 13 (0.22)&Khaligh et al.~\cite{Khaligh2009}& C & -\\\hline 
6 (0.67)&  - (0)&70&1, 2, 4 (1.5)&14, 15 (0.11)&Sharma and Baredar~\cite{sharma2019}& C & Depolarization, sudden breaking of piezo layer due to high brittleness\\\hline 
4 (0.44)&  100$\mu W$ (1)&33&4 (0.5)&1, 2, 3, 4, 5, 6 (0.33)&Mateu and Moll~\cite{Mateu2005}& C & A discussion on power consumption of microelectronic devices\\\hline 
4 (0.44)& - (0) &153&1, 2, 4 (1.5)&6, 11, 14, 15 (0.22)&Calio et al.~\cite{calio2014}& C & Optimal shapes, buckling\\\hline 
1 (0.11)&  ~$\mu W$ to $mW$ (1) &95&4 (0.5)&11, 12, 14, 15, 16 (0.28)&Batra et al.~\cite{Batra2016}& D & -\\\hline 
3 (0.33)&  1.3$mW$ (1)&16&4 (0.5)&-(0)& Sun et al.~\cite{Sun2012} & D & Comparison with energy from wind, solar, geothermal, coal, oil and gas\\\hline
1 (0.11) & - (0)&13&-(0)&14, 15, 16 (0.17) & Sharma et al.~\cite{Sharm2018} & E & Historical points\\\hline
\hline
\end{tabular}
}
\label{general11}
\end{table*}

Although most of the aforementioned general review papers have more or less similar titles, but their scientific depth and the number of reviewed items are different. 
Some papers like Ref.~\cite{calio2014} have focused on design strategies of the piezoelectric energy harvesters. They try to present a guide for the selection of piezoelectric materials as harvesters. Moreover, almost all the mentioned reviews suffer from weak classifications stemming from generality of their topic.

The results of evaluation of generally-written review papers on piezoelectric energy harvesting have been presented in Table~\ref{general11}. The table contains different sub-categories, the range of output power, the number of reviewed articles, the merits, general conclusions, and some other extra descriptions. The grade for each paper has been computed based on the number of merits, the number of subcategories, the number of concluding remarks, and declaration of minimum required output power.

\section{Design and fabrication}
\subsection{Materials}
The choice of suitable piezoelectric material is a critical step in designing energy harvesters~\cite{Gos2019}. Thus, lots of the review papers in the field of energy harvesters less or more have addressed the piezoelectric materials. Different performance metrics have been selected for comparing piezoelectric materials on diverse applications. In actuating and sensing applications, the piezoelectric strain and piezoelectric voltage constants are appropriate criteria. However, the electromechanical coupling factor, power density, mechanical stiffness, mechanical strength, manufacturability, and quality factor are the most important factors for energy harvesting. Also the operating temperature is important  in material selection ~\cite{Bed2009}.

Li et al.~\cite{Li2014} divided the piezoelectric materials into four categories (ceramics, single crystals, polymers, and composites) based on their structure characteristics. They described the general properties of these four piezo-material categories, and compared some of the most important candidate materials form these categories in terms of piezoelectric strain constant, piezoelectric voltage constant g, electromechanical coupling factor k ,mechanical quality factor Q , and dielectric constant e.  They commented that  piezoelectric ceramics and single crystals  have much better piezoelectric properties than piezoelectric polymers that is due to the strong polarizations in their crystalline structures. On the other hand, piezoelectric ceramics and single crystals are more rigid and brittle then piezoelectric polymers. Both piezoelectric properties and mechanical properties are important in selection of a certain piezoelectric material for a specific piezoelectric harvesting application. Other important parameters in selecting the suitable materials are the application frequency, the available volume, and the form in which mechanical energy is fed into the system. In order to harvest maximum amount of energy, the piezoelectric energy harvester should operate at its resonance frequency. However, in many cases such as low frequency applications,  it is impractical to match the resonance frequency of the piezoelectric with the input frequency of the host structure. They demonstrated that for low frequency applications in off-resonance conditions the piezoelectric element can be approximated as a parallel plate capacitor and for harvesting more electric energy the product of piezoelectric strain constant and piezoelectric voltage constant should be high. On the other hand, for near-resonance conditions, the optimum output power of the harvester is independent of piezoelectric properties of piezo-element but the maximum output voltage depends on piezoelectric strain constant. It is obvious that the selection of suitable piezomaterial for piezo-harvester depends on working condition, and it makes the selection of piezo-material more complex. The article has not specified the minimum required power output for piezoelectric energy harvesters. Also, the energy density of piezoelectric materials were not reported. The focus is on macroscale piezomaterials and micro- and nono-scale materials were not covered.

\begin{figure}[htbp] \setlength{\unitlength}{1mm}
\begin{picture}(80,50)
\includegraphics[height=50mm]{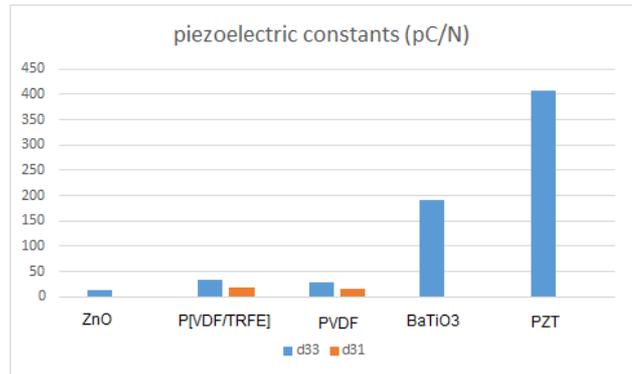}
\end{picture}
\caption{Piezoelectric coefficient range for some of piezoelectric materials~\cite{Narita2018}.}
\label{Fig:A}
\end{figure}

Narita and Fox~\cite{Narita2018} reviewed three categories including piezoelectric ceramics/polymers, magnetostrictive alloys, and magnetoelectric multiferroic composites. Their review included describing the properties of PZT, PVDF, ZnO. They compared some of the piezoelectric materials based on their piezoelectric coefficients (Fig.~\ref{Fig:A}). Also, they remarked some advantages and disadvantages of traditional piezoelectric ceramics, piezoelectric polymers, and composites. They focused on characterization, fabrication, modeling, simulation, durability and reliability of piezo-devices. Based on their analysis, the future directions include the device size reduction to make them suitable for nanotechnology, optimization, and developing accurate multi-scale computational methods to link atomic, domain, grain, and macroscale behaviors. Investigation of temperature-dependent properties, development of materials and structures capable of withstanding prolonged cyclic loading, duration of electro-magneto-mechanical properties, and fracture/fatigue studies are other recommendations for future research. The review does not reported some of the important mechanical and piezoelectric properties of the piezo-materials like electromechanical coupling factor and quality factor, mechanical strength and mechanical stiffness, and the materials were compared based on their piezoelectric coefficients and the output power of the energy harvesters.

Safaei et al. reviewed the recent progresses in the field of piezoelectric ceramics like soft and hard PZTs, piezoelectric polymers including PVDF, piezoelectric single crystals, lead-free piezoelectrics, high temperature piezoelectrics, piezoelectric nanocomposites, and also piezoelectric foams. They reported the piezoelectric coefficient, and the maximum output voltage for some of these materials without describing the geometry of the piezoelectric harvester. Brittleness of PZTs and existence of health risks in PZT ceramics due to the toxicity of lead are the most important challenges of using PZTs, which motivates the development of lead-free flexible and high-performance piezoelectric materials. They concluded that the need for enhancement of electromechanical, thermal, and biocompatible properties has led to the introduction of new piezoelectric materials including new lead-free piezoelectrics, high-temperature piezoelectrics, piezoelectric foams, and piezoelectric nanocomposites. The paper have eplained lots of piezo-materials however there is not a systematic comparison between the piezoelectric materials in terms of piezoelectric and mechanical properties. It seems that the main target is only reporting the recent progresses in the field. Also the minimum required output power for the piezoelectric harvesters  was not remarked.

Zaarour et al.~\cite{zaar2019} summarized the energy harvesting technologies developed based on piezoelectric polymeric fibers, inorganic piezoelectric fibers, and inorganic nanowire. The paper contains a review of piezoelectric fibers and nanowires with respect to the peak voltage, the peak current, the active area, and their advantages, without describing the working conditions and mechanical structure of the related piezoelectric energy harvester. Maybe due to lack of available data on properties of nano-scale piezoelectric materials, there is not any comparison between the selected materials in terms of their piezoelectric and mechanical properties. The reported output powers are in the range of micro watt which is not enough for empowering real electronic systems and circuits. They concluded that, standardizing the performance of the piezo-nanogenerator, developing effective packaging technology, packaging of nano-piezo-harvesters, commercializing products for harsh environments, finding a suitable approach to enhance the electrical outputs, and augmenting the durability and the output stability are some future horizons.

Yuan et al.~\cite{Yuan2016} introduced the dielectric electroactive polymers as promising replacements for conventional piezoelectric materials. Electroactive polymers are lightweight, flexible, ductile, low-cost manufactured, with high strength-to-weight ratio, low mechanical impedance, and can endure large strains. The dielectric polymers need high voltage to realize energy cycles that may lead to the breakdown of the device. Piezoelectric materials are employed in energy harvesters because of their compact configuration and compatibility. However, these materials have inherent limitations including aging, depolarization, and brittleness. In comparison, electrostrictive polymers are promising candidates to replace piezoelectric materials in vibration energy harvesting cases. The challenge in design of electroactive polymer energy harvesters is to develop systems that are capable of ensuring a constant initial voltage on the polymer at small cost.

There are some other review papers which have focused on several issues in the field of piezoelectric materials.  Piezoelectric polymers were reviewed some of papers like Mishara's review paper et al.~\cite{Mishra2019}. High temperature single crystals is the subject of Priya's  paper which made a comparative study of the main high temperature piezoelectric single crystals. Bio-piezoelectric materials were described by Liu et al.~\cite{Liu2018}. Also they have reviewed micro and nano fabrication techniques for micro/nano scale energy harvesters. useful information on micro/nano scale piezoelectric materials may be found in Gosavi et al. ~\cite{Gos2019}. They defined a systematic roadmap to select the piezoelectric materials for micro and nanoscale energy harvesters. They pointed out that the ZnO thin film is the most widely used structure in micro and nanoscale harvesters, and can be economically synthesized in arbitrary sizes and shapes. A detailed comparison between traditional macro materials and new micro/nano piezoelectric materials in terms of dielectric, mechanical and piezoelectric properties was performed by Bowen et al.~\cite{bown2014}. They mentioned some points about high-temperature harvesting related to the Curie temperature, light harvesting into chemical or electrical energy, and optimization algorithms. Their investigation contains parameters like pyroelectric coefficient (harvesting from temperature fluctuations), the electro-mechanical coupling, the mechanical quality factor, the constant-strain relative permittivity, the constant-stress relative permittivity, the piezoelectric coefficient, and the elastic constant of piezoelectric materials. For high-strain applications, they suggested polymeric or composite-based systems. Their suggested future directions are understating and development of new materials and gaining strong scientific underpinning of the technology and reliable measurements.

Most of the review papers tried to compare the piezoelectric materials and draw a roadmap for selecting an appropriate material for energy harvesters. However, the choice of material is strictly dependent on type of the energy harvester, its working condition, the cost level, accessibility and ease of fabrication/synthesis of the piezoelectric material. For example, Ullah Khan and Ahmad~\cite{KJhan2016} who have reviewed vibrational energy harvesters utilizing bridge oscillations, pointed out that the main selection criteria for piezoelectric vibrational energy harvesting are the dielectric constant, the Curie temperature, and the modulus of elasticity of the material.

The piezoelectric materials with high value of elastic modulus can be an appropriate choice for high acceleration vibrations. However, the piezoelectric materials like lead lanthanum zirconate titanate that has a high value of dielectric constant will perform very well in low-acceleration vibrational environments. Also, due to the easiness of in situ fabrication of lead zirconate titanate (PZT) with sol-gel technique, and its easy integration with the other microfabrication processes, PZT has been largely utilized in most of such applications.

As another example, we can point out the selection of a desirable piezoelectric material for walking energy harvesting applications. Based on a review performed by Maghsoudi Nia et al.~\cite{nia2017}, this application needs an incombustible, chemically resistant, low-price material, which should be unbreakable under harsh conditions. The mentioned criteria have made PVDF more suitable than PZT for the most of piezoelectric harnessing from walking. Most of the review papers have contented themselves with reporting some electromechanical properties of piezoelectric materials, and provided a few information on the accessibility, relative cost, chemical properties, ease of fabrication, and suitable working conditions of different piezoelectrics. The lack of such information indicates the need for further research and also the necessity for making more comprehensive and application-based reviews on piezoelectric materials.

The results of evaluation of the review papers on piezoelectric materials have been presented in Table~\ref{matter11}. The table also contains different sub-categories, the range of output power, the number of reviewed articles, the merits, general conclusions, and some other extra descriptions. The rank of each paper has been computed based on the number of merits, the number of subcategories, the number of concluding remarks, and clear emphasizing on value of minimum required output power.
As indicated by table 1 unless a few papers like  ~\cite{Li2014} other reviews suffer from lack of reported data on mechanical piezoelectric materials, their fabrication methods and other figure of merits in material selection. Also, unless the paper ~\cite{Narita2018} which has pointed out the needed energy for empowering the electronic devices, other papers have neglected the minimum required energy for an energy harvester.

\begin{landscape}
\begin{table}
\caption{{\normalsize Overall evaluation of review papers written on materials in piezoelectric energy harvesting. The numbers in brackets denote non-general future lines. "Cons." stands for conclusions. \\
\textbf{Conclusions:} Efficiency/performance improvement, cost reduction, lead free materials, increasing life time, endurance, size reduction and manufacturability. \\
\textbf{Merits:} 1: piezoelectric coefficients, 2: coupling factors, 3: manufacturability, 4: mechanical strength, 5: guidelines for material selection, 6: paying attention to the minimum required output power (1 mW), 7: energy density, 8: stiffness, 9: quality factor \\
\textbf{Sub-categories:} 1- Micro and nano materials: 1-1 Piezoelectric micro/macro fibers, 1-2 Polymer nano fibers, 1-3 Ceramic nano-fibers, 1-4 Piezoelectric nano wires, 1-5 Micro/nano fibers/wires composites.\\
2- macro scale materials: 2-1- Piezoelectric polymers (PVDF, Pu, P(VDF-TrFE), cellular PP), 2-2-piezoelectric ceramics (PZT, PMM-PT, PMN-PZT ...), 2-3- piezoelectric single crystals (Quartz ...), 2-4- piezoelectric foams (PDMS piezoelectric, PET/EVA/PET piezoelectret, FEP piezoelectric), 2-5- piezoelectric powders, 2-6- piezoelectric composites (PVDF with Nanofillers, Non-Piezoelectric Polymer with BaTiO3), 2-7-bio materials.
}
}
\resizebox{1.25 \textwidth}{!}{
\begin{tabular}{c|p{2.5cm}|p{1.0cm}|p{3.0cm}|p{2.2cm}|p{2.7cm}|p{0.8cm}|p{15.5cm}} \hline\hline
\# Cons. & Minimum required output & \# Refs. & Merits & Sub-categories & Ref. & Grade & Highlights \\ \hline\hline
6 (0.75)& ~$\mu W$ to $mW$ (0) & 120 &1, 2, 3, 5, 6, 8, 9 (1.75) &2-1, 2-2, 2-3, 2-6 (0.33) & Li et al.~\cite{Li2014} & C & 1 the current state of research on piezoelectric energy harvesting devices for low frequency (0-100 Hz) applications and the methods that have been developed to improve the power outputs of the piezoelectric energy harvesters have been reviewed. 2 The selection of the appropriate piezoelectric material for a specific application and methods to optimize the design of the piezoelectric energy harvester were discussed. \\
6 (0.75) & ~$\mu W$ to $nW$ (0) & 478 &1, 3, 6, 7 (1.15) & 1, 2-2, 2-3 2-4, 2-6 (0.75)&Safae et al.~\cite{Safaei2019}& C &Reporting the recent advances in the field of piezoelectric materials. Reviewing some novel piezoelectric materials like piezoelectric foam and high temperature materials  \\
9 (1) &~$\mu W$ (0) & 173 &3, 4, 5 (0.75) &1-1 to 1-5 (0.42) &Zaarour et al.~\cite{zaar2019}& C & 1 Manufacturing methods of nano fibers and wires, 2 Mentioning output voltage and currents of nano/micro materials, 3 Comparison of nano/micro materials based on maximum voltage and currant and active area\\
3 (0.4)&~$\mu W$ (0) & 446 &1, 3, 5 (0.75) &1-2, 1-3, 2-1, 2-2, 2-3, 2-4 (0.35) &Liu et al.~\cite{Liu2018}& D & 1 Reporting recent progresses in the field of piezoelectric materials, 2- description of fabrication techniques of lots of piezoelectric materials in energy harvesting applications, 3- explaining the main frequency bandwidth broadening techniques, 4- Classifying piezoelectric materials, fabrication techniques, and frequency bandwidth broadening techniques.\\
6 (0.75)&~$\mu W$ to $mW$ (0) & 175 &1, 3, 6 (0.75) &1-1, 2-1, 2-2, 2-3 (0.33) &Narita and Fox~\cite{Narita2018} &  D & 1 Reporting the harvested power of PZT based PEH s with different structures, 2 Reporting the recent advances in the field op PEHs which were made of PVDF, and polymer based composite piezoelectrics. Comparing the output power of some of the piezoelectric energy harvesters. \\
2 (0.25) & ~$mW$  & 50 & 1, 4, 7, 8 (1.15) & 1, 2 (0.25) & Yoan et al.~\cite{Yuan2016} &  D &1 introducing electrostrictive and dielectric electro-active polymers, 2 performance comparison of PZT, PVDF, and DEAPs and electrostrictive polymers. Describing the industrial challenges for dielectric electro-active polymers.\\
4 (0.57)&~$\mu W$ (0)& 158 &1, 2, 3 (0.75) & 2-1, 2-2, 2-3, 2-6 (0.33) & Mishra et l.~\cite{Mishra2019}& D &The article basically aimed at exploring the basic theory behind the piezoelectric behavior of polymeric and composite systems and comparing the important types of piezoelectric polymers and composites. The article described the piezoelectric properties of lots of the piezo-polymers and polymer composites. \\
6 (0.75)&~$\mu W$ (0) & 216 &1, 2, 9 (0.75) &2-1, 2-2, 2-3 (0.25)& Bowen et al.~\cite{bown2014}& D & Reviewing some resent topics like piezoelectric light harvesting, Pyroelectric based harvesting, and nano scale Pyroelectric systems\\
3 (0.4)&~$\mu W$ (0) & 24 &2, 6, 9 (0.75) &2-1, 2-3 (0.25) &Lefeuvre~\cite{Lef2009}& D & 1 Figure of merit for energy conversion efficiency, 2 figure of merit for piezoelectric materials, 3 comparing the one, two and three stage electric power interfaces \\
2 (0.25)&~$\mu W$ to $mW$ (0) & 16 & 1, 2, 5, 8 (1) & 2 (0.1) & Mukherjee and Datta~\cite{Mukh2010}& D & 1 Effect of load resistance on the output power of PEHs, 2 Selection criteria for piezoelectric ceramics \\
\hline\hline
\end{tabular}
}
\label{matter11}
\end{table}
\end{landscape}

\subsection{Structure}
All piezoelectric energy harvesters include a mechanical part (or transduction part) to convert the input mechanical energy into the electric charges in the piezoelectric element, and an electric part that keeps the electric charges and converts them into a suitable form of electric output like direct voltage. Design of the mechanical part of a piezoelectric energy harvester usually includes the determination of its size, configuration, working modes, and selection of appropriate materials to enhance its performance characteristics like the output electric energy, the conversion efficiency and the working bandwidth. The size of the piezoelectric energy harvester may vary from micro and nanoscale (lower than 0.01$cm^3$) to macroscale (75$cm^3$) ~\cite{Dut2005}.

Upon the literature, the piezoelectric energy harvesters can be classified from different viewpoints. Form the view point of operating frequency, they may be categorized into two main sections: the resonant type devices that operate at or near their resonance frequency, and non-resonant systems that do not depend on any specific frequency. The piezoelectric energy harvesters may harvest energy from motions in a unique direction or from multi-directions. Accordingly, they may be single-directional or multi-directional harvesters. Also, they have a single or several vibration modes (multi-modal harvesters). From the viewpoint of governing dynamic models, the piezoelectric harvesters may be linear or non-linear~\cite{Maar2019}. As indicated in Fig.~\ref{Fig:map}, their configuration can be classified as cantilever type, stack type, cymbal type, circular diaphragm type, or the shell and film types.

\begin{figure}[htbp] \setlength{\unitlength}{1mm}
\begin{picture}(80,50)
\includegraphics[height=50mm]{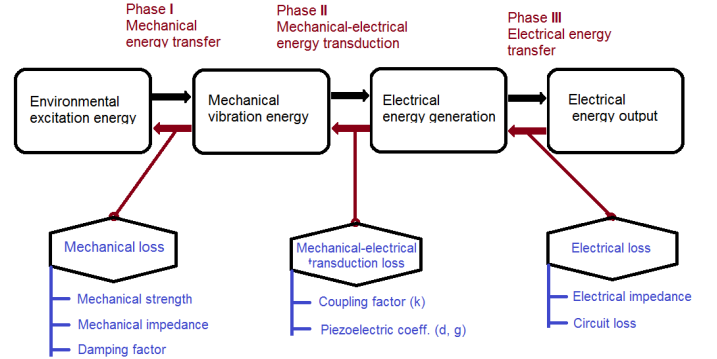}
\end{picture}
\caption {Three major phases associated with piezoelectric energy harvesting~\cite{Uch2018}.}
\label{Fig:B}
\end{figure}

Uchino~\cite{Uch2018} started his review by mentioning the historical background of the piezoelectric energy harvesting, and explaining several important misconceptions. He reviewed the different design approaches followed by mechanical, electrical, and MEMS engineers. He remarked that there are three major phases associated with piezoelectric energy harvesting: (i) mechanical-mechanical energy transfer, (ii) mechanical-electrical energy transduction, and (iii) electrical-electrical energy transfer to accumulate the energy into a rechargeable battery. Fig.~\ref{Fig:B} represents these three major phases. In order to provide comprehensive strategies on how to improve the efficiency of the harvesting system, a step-by-step detailed energy flow analysis is essential. It was mentioned that the five important figure of merits in piezoelectrics are the piezoelectric strain constant $d$, the piezoelectric voltage constant $g$, the electromechanical coupling factor $k$, the mechanical quality factor $Q_m$, and the acoustic impedance $Z$. Also, the energy transfer rates for the piezoelectric energy harvesting systems with typical stiff cymbals and flexible piezoelectric transducers were evaluated for three aforementioned phases/steps. Moreover, a hybrid energy harvesting device that operates under either magnetic and/or mechanical noises was introduced. It was concluded that the remote signal transmission, energy accumulation in rechargeable batteries, discovering a genius idea to combine nano-devices in parallel, and enhancing the energy density in medical applications have been introduced as future research fields. It was declared that a clear future perspective for NEMS and MEMS piezoelectric harvesters is missing due to their low energy levels (in the order of pW to nW). We need to discover a genius idea on how to combine thousands of nano-devices in parallel and synchronously in phase. Describing the performance improvement techniques for non-resonant and resonant energy harvesters, are felt missing in this article.

\begin{figure}[htbp] \setlength{\unitlength}{1mm}
\begin{picture}(80,60)
\includegraphics[height=60mm]{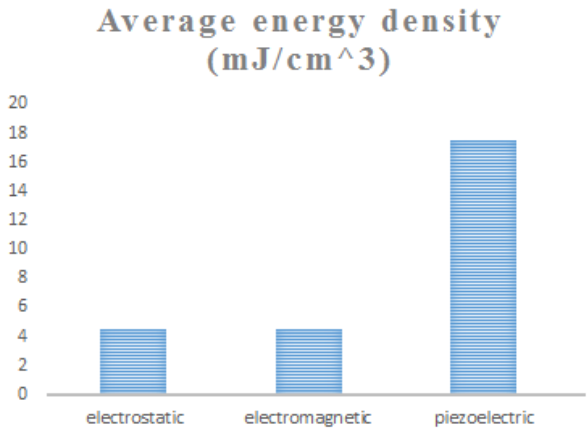}
\end{picture}
\caption {Comparison of the energy density for the three types of mechanical to electrical energy converters~\cite{priya2007}.}
\label{Fig:C}
\end{figure}

Priya~\cite{priya2007} classified the energy harvesting approaches in two categories (1) power harvesting for sensor networks using MEMS/thin/thick film approach, and (2) power harvesting for electronic devices using bulk approach. His review article covered the later category in more details. He listed almost all the energy sources available in the surrounding which may be used for energy harvesting and commented that the selection of the energy harvester as compared to other alternatives such as battery depends
on two main factors cost effectiveness and reliability. Also, he reported the daily average power consumption for a wearable device, and of common household devices. Next, comparison of the energy density for the three types of mechanical to electrical energy converters including electrostatic, electromagnetic and piezoelectric were performed. The results were represented in Fig.~\ref{Fig:C}. He concluded that piezoelectric converters are prominent choice for mechanical to electric energy conversion because the energy density is three times higher as compared to electrostatic and electromagnetics. He gave a review of piezo-harvesters appropriate for light-weight flexible systems with easy mounting, large response, and low-frequency operation; called the low-profile piezo-transducer in on/off-resonance condition. A good discussion on piezoelectric polymers, energy storage circuit, and microscale piezo-harvesting device is available in the article. He mentioned that the electrical power generated by the piezoelectric energy harvester is inversely proportional to the damping ratio that should be minimized through proper selection of the material and design. He also have summarized the conditions leading to appearance of maximum efficiency in low profile piezoelectric energy harvesters. An interesting part of the paper is the description of the piezoelectric material selection procedure for on/off-resonance condition. However, the description of performance improvement techniques for enhancing the system frequency response are felt missing in this article.

Yang et al.~\cite{Yang2018} commented that from the perspective of applications,the output power of the harvester and its operational frequency bandwidth are the two metrics most useful to product development engineers. They explained the materials selection procedure for piezoelectric energy harvesters in off-resonant condition and remarked why PZT's are steel the most popular piezoelectric materials for energy harvesters. They stated that linear resonant harvesters are not suitable for harvesting energy
from broadband or frequency-varying excitations, and in this condition  nonlinear energy harvesters have been proven to be able to exhibit a broadband performance. Therefore, researchers have explored monostable, bistable, and tristable systems and developed some frequency tuning approaches, such as multi-cantilever structures, bistable composite plate designs, and passive and active stiffness-tuning technologies. On the nonlinear energy harvesters they remarked that maintaining the nonlinear harvesters in the high-energy oscillation states, especially under weak excitations is a difficult task. specially, with zero initial conditions, nonlinear harvesters usually follow the low-energy orbits, which results in small-amplitude voltage responses. Thus maintaining the nonlinear PHE in the high-energy states is a critical problem which is possible with active and passive control. Efficiently transferring and storing the generated broadband or random electric energy is another critical problem for nonlinear PHEs. Moreover, they reviewed the different designs strategies, the optimization techniques, and the harvesting piezo-materials in applications like shoes, pacemakers, tire pressure monitoring systems, bridge and building monitoring. They declared that high energy conversion efficiency, ease of implementation, and miniaturization are the main advantages of such systems. However, authors state that enhancement of energy efficiency of the piezo-based harvesters is still an open challenge. They also made a systematic performance comparison on some of the energy harvesters. They pointed out that a considerable gap exists between the achieved performance and the expected performance. Therefore, in situ testing, applying more realistic excitations, system-level investigations on piezo-harvesters integrated with the power conditioning circuits, energy storage elements, sensors, and control circuits need to be investigated.  This article has focused on mechanical part of energy harvesters and subjects like the electric interface circuits of the harvesters and their energy flow analysis have not been remarked.

There are some other review papers which have focused on several issues in the field of design of piezoelectric harvesters. Performance improvement techniques for PHEs and design optimization methods are hot topics covered by reviews ~\cite{yild2017},~\cite{Li2014},~\cite{Maar2019} . Manual and autonomous tuning systems for widening the operating frequency bandwidth and the future plans in this field were discussed by Ibrahim and Vahied~\cite{Ibrahim2012}.  A good review of PEH configurations such as cantilever beam, discs, cymbals, diaphragms, circular diaphragms, shell-type, and ribbon geometries may be found in ~\cite{Li2014}. Talib et al.~\cite{Tal2019} explained effective strategies and the key factors to enhance the performance of piezoelectric energy harvesters operating at low frequencies, including selection of the piezoelectric material, optimization of the shape, size, structure, and development of multi-modal, nonlinear, multi-directional, and hybrid energy harvesting systems. This review paper is suitable for the beginners who want to get acquainted with the piezoelectric materials and some designs of piezoelectric energy harvesters. They concluded that the recent developments are inclined towards generation of more power from low-frequency and low-amplitude ambient vibrations with reduced required piezoelectric material. Adding a single DOF system in the form of an extension beam or a spring to the piezoelectric beam is a remarkable advise to enhance the power output. They showed that the multi-modal energy harvester exhibits a broader bandwidth when its multiple resonance peaks get closer.

Brenes et al.~\cite{Bren2020} provided an overlook of existing energy harvesting circuits and techniques for piezoelectric energy scavenging to distinguish between existing similar solutions that are different in practice. Such categorization is helpful to ponder the advantages and drawbacks of each available item. Their review is unique since they have classified the piezo-systems based on adaptive/non-adaptive control strategies, topologies, architectures, techniques form one hand, and electromechanical models from the other hand. The best system has been introduced with respect to the optimized power efficiency, the design complexity, the strength of coupling, the multi-stage load adaption, and the vibration frequency.

Issues like AC-DC conversion mechanism, the passive and active rectifications, the start-up issues, the harvester-specific interactions, the voltage conditioning, the DC-DC charge pumps, the power regulation, and the impedance matching were discussed by Szarka et al.~\cite{Szarka2011} and Dell'Anna et al.~\cite{Dell2018} . Non-linear electronic interfaces for energy harvesting from mechanical vibrations was remarked in ~\cite{Guy2011}.

Tables~\ref{design} and~\ref{power} gather together the results of evaluation of the review papers written on design methods and the power interface considerations, respectively. The table also contains different sub-categories, the range of output power, the number of reviewed articles, the merits, general conclusions, and some other extra descriptions. The rank of each paper has been computed based on the number of merits, the number of subcategories,  the number of concluding remarks, and clear emphasizing on value of minimum required output power.

The results of evaluation of review papers on design of piezoelectric energy harvesting have been presented in Table~\ref{design}. The table also contains different sub-categories, the range of output power, the number of reviewed articles, the merits, general conclusions, and some other extra descriptions. The grade for each paper has been computed based on the number of merits, the number of subcategories, the number of concluding remarks, and declaration of minimum required output power.
Table~\ref{design} is designed to evaluate the review papers about design of PHEs. The merits that are selected as the necessary considerations in the field of design of PHEs are 1: reporting the output power of PHEs, 2: reporting the coupling factors and operational modes, 3: including mathematical models, 4: attending to the motivating frequencies of PHEs, 5: Attending to the mechanical and electrical energy conversion efficiencies. Quantitative evaluation of the papers was performed based on the number of merits which have been followed by the article, number of sub-categories which were covered in the review and the number of conclusions. according to the table, unless the first papers, other papers have neglected some merits like the minimum required output power for the harvesters and energy flows analysis of them, also most of the papers have not reviewed some of the issues in the field as reported by paper "*".

\begin{landscape}
\begin{table}
\caption{ Overall evaluation of review papers written on design of piezoelectric energy harvesters. The numbers in brackets denote the number of non-general future lines. "Cons." stands for conclusions. \\
\textbf{Conclusions:} Efficiency/performance improvement, necessity of frequency bandwidth broadening, necessity of optimizations, increasing life time and endurance, size reduction and manufacturability, importance of electric interface circuits, the necessity of material properties improvement. \\
\textbf{Merits:} 1: reporting the output power of PHEs, 2: coupling factors and operational mode, 3: including mathematical models, 4: matching the resonance frequency of PHEs with motivating frequencies, 5: paying attention to the energy conversion efficiencies \\
\textbf{Sub-categories:} Performance improvement: 1- frequency tuning approaches: 1-1-manual tuning and 1-2-autonomous tuning methods, 2- Multi-frequency systems, 3- nonlinear systems, 4- frequency up conversion approach, 5- systems with
free moving mass , 6- Bi directional and three directional systems, 7- Amplification techniques 8- material selection
criteria 9- energy conversion efficiency 10- low profile piezoelectric harvesters 11- geometric optimization 12- Mathematical modeling of PHEs. Design improvements for 13- piezoelectric cantilevers, 14- piezoelectric cymbal 15- piezoelectric stack configuration, 16- electrode optimization.... 17-performance quantification and comparison strategies...18- electronic interface circuits for PHEs 19- Hybrid energy harvesting mechanism.
}
\resizebox{1.25 \textwidth}{!}{
\begin{tabular}{c|p{2.5cm}|p{1.2cm}|p{2.5cm}|p{2.7cm}|p{2.2cm}|p{0.8cm}|p{15.5cm}} \hline\hline
\# Cons. & Minimum required output & \# Refs. & Merits & Sub-categories & Ref. & Grade & Highlights \\ \hline\hline
10(1)&$mW$ (1)&35&1, 2, 3, 4, 5 (2) &8, 9, 11, 13, 14, 15, 17, 18, 19 (0.5) &Uchino~\cite{Uch2018}& A & 1 describing the historical background of piezoelectric energy harvesting, 2 commenting on several misconceptions by the current researchers, 3 step-by-step detailed energy flow analysis in energy harvesting systems, 4- describing the key to dramatic enhancement in the efficiency, 5- important comments on the useful/un-useful output power level for the harvesters\\
6(0.85)&$mW$ (1)&75&1, 2, 3, 4, 5 (2) &9, 8, 10, 12 (0.2) & Priya~\cite{priya2007} & A &Describing the material selection criteria in on- and off-resonance condition, Describing the factors which affect the conversion efficiency of PHEs, introduction of some low profile PHEs for realizing a self powered sensor nodes\\
5(0.7)&~$\mu W$ to $mW$ (1)&338&1, 2, 3, 4, 5 (2) &3, 11, 12, 13, 14, 15, 16, 17 (0.36) &Yang et al.~\cite{Yang2018}& A & Analysis of different designs, nonlinear methods, optimization techniques, and materials for increasing performance. Introducing a set of metrics for the end users of PHEs for comparison of performance of PHEs \\
5(0.7)&~$\mu W$ to $mW$ (0) & 120 &1, 2, 3, 4, 5 (2) &1, 4, 8, 12, 14, 18 (0.32) &Li et al.~\cite{Li2014}&B & Commenting on the biggest challenges for PHEs, describing the most important limitations of piezoelectric materials\\
4(0.75)&~$\mu W$ to $mW$ (0) &446 &1, 2, 3, 4, 5 (2) &2, 3, 4, 6, 12, 13, 14, 15, 19 (0.52) &Liu et al.~\cite{Liu2018}& B &Various key aspects to improve the overall performance of a PEH device are discussed. Classification of performance improvement approaches have been performed.\\
3(0.42)&~$\mu W$ (0) &149  & 1, 3, 4, 5 (1.6) & 1, 2, 3, 4, 5, 6 (0.3) & Maamer et al.~\cite{Maar2019}& C
&Proposing new generic categorization, approach based on the improvement aspect of the harvester, which includes techniques for widening operating frequency, conceiving a non-resonant system and multidirectional harvester. Evaluating the applicability of the performance improvement techniques under different conditions and their compatibility with MEMS technology \\
5(0.7) &$mW$ (0) &105 &1, 3, 4 (1.2) &1, 2, 3, 4, 5, 6, 7 (0.36) &Yildirim et al.~\cite{yild2017}& C & New classification of performance enhancement techniques, Comparison of lots of performance enhancement techniques.\\
8(1) &~$\mu W$ to $mW (0)$ & 66 & 1, 3, 4 (1.2) & 1-1, 1-2 (0.1) &Ibrahim and Vahied.~\cite{Ibrahim2012}& C
& Classifying, reviewing and comparing the different manual and autonomous tuning methods, challenge of energy consumption by self-tuning structures\\
4(0.6)&~$\mu W$ to $mW$ (0) & 135 &1, 2, 4, 5 (1.6) &3, 6, 8, 11 (0.25) & Talib et al.~\cite{Tal2019}& C &They commented that the anticipated performance of a piezoelectric harvester can be attained by achieving the trade-off between output power and bandwidth.\\
\hline\hline
\end{tabular}
}
\label{design}
\end{table}
\end{landscape}

\begin{table*}
\caption{{\normalsize Overall evaluation of review papers written on power interfaces in piezoelectric energy harvesters. The numbers in brackets denote the number of non-general future lines. "Cons." stands for conclusions. \\
\textbf{Conclusions:} Efficiency/performance improvement, necessity to considering interactions between the mechanical harvester and the power electronics, necessity of optimizations, importance of electric interface circuits \\
\textbf{Merits:} 1: Energy flow analysis of PHEs, 2: Practical implementation of electronic interfaces, 3: including mathematical models, 4: paying attention to electrical impedance matching, 5: paying attention to the energy consumption of electric interfaces, 6: analysis of energy conversion efficiency. \\
\textbf{Sub-categories:} 1: Three Phases in Energy Harvesting Process, 2: mechanical-electrical energy transduction, 3: energy flow analysis, 4: electrical-to-electrical energy transfer, 5: DC-DC converters and conversions, 6: electric impedance matching, 7: electromechanical models of PHEs, 8: requirements for power electronics, 9: AC-DC conversion with voltage conditioning, 10: DC-DC conversion with voltage conditioning, 11: Power regulation, 12: conversion efficiency of PHEs, 13: rectification approaches: (13-1 resonant PEH rectifiers, 13-2- series synchronized switch harvesting on inductor (S-SSHI), 13-3- synchronized switching and discharging to a storage capacitor through an inductor (SSDCI) rectifier, 13-4- synchronous electric charge extraction (SECE), 13-5- synchronized switch harvesting on inductor magnetic rectifier (MR-SSHI), 13-6- hybrid SSHI, 13-7- adaptive synchronized switch harvesting (ASSH), 13-8- enhanced synchronized switch harvesting (ESSH), 13-9 MPPT-based PEH Rectifiers), 14: performance of rectification approaches, 15: autonomous switch control in resonant PEH rectifiers, 16: switching techniques, 17: parallel SSHI, 18: load decoupling interfaces, 19: non-adaptive MPPT control, 21: characteristics of existing adaptive control strategies, 20: the tunable OSECE technique, 21: two-stage load adaptation FB technique, 22: two-stage load adaptation shunt rectifier technique, 23: PS SECE technique, 24: tunable SECE technique 25: tunable USECE technique, 26: N-SECE technique, 27: FTSECE technique, 28: HB and FB 3-stage load adaptation technique, 29: tunable SCSECE technique, 30: four-stage topology: the SSH architecture, 31: parallel SSHI (p-SSHI) technique, 32: Series SSHI (s-SSHI), DSSH and ESSH techniques, 33: technical guidelines for the choice of an adequate circuit.}
}
\resizebox{1.0 \textwidth}{!}{
\begin{tabular}{c|p{2.5cm}|p{1.2cm}|p{1.5cm}|p{1.5cm}|p{1.7cm}|p{1.0cm}|p{7.0cm}} \hline\hline
\# Cons. & Minimum required output & \# Refs. & Merits & Sub-categories & Ref. & Grade & Highlights \\ \hline\hline
9(1)& $mW$ (1)&35 &1, 2, 3, 4, 5, 6 (2) & 1, 2, 3, 4, 5, 6, 7, 8, 12 (0.27) & Uchino~\cite{Uch2018} & A& Mentioning minimum acceptable output power for harvesters, energy flow analysis for cymbal type transducer, describing the electric impedance matching technique \\
3(0.5) &- (0) &109 &1, 2, 3, 4, 5, 6 (2) & 12 to 34 (1) & Brenes et al.~\cite{Bren2020} & B & Comparison of the conditions for electric tuning techniques to maximize the power flow from an external vibration source to an electrical load description of necessary conditions for Maximum Power Point Tracking (MPPT) \\
8(1) & ~$\mu W$ (0) & 113 &2, 3, 4, 5, 6 (1.66) & 1, 2, 4, 5, 8, 9, 10, 11, 13 (0.27) & Szarka et al.~\cite{Szarka2011} & B & Overview of power management techniques that aim to maximize the extracted, power of PHEs Describing the Requirements for power electronics reviewing various power conditioning techniques and comparing them in terms of complexity, efficiency, quiescent power consumption, startup behavior \\
6(0.75)&-(0) &113 & 2, 3, 4, 5, 6 (1.66) &7, 12, 13 (all items), 14, 15 (0.15) & Francesco et al.~\cite{Dell2018}& C
&1: Almost all the rectification techniques employed in PEH systems were discussed and compared emphasizing the advantages and disadvantages of each approach. 2: Introducing the seven criteria used to evaluate the performance of a harvesting interface \\
1(0.2) &- (0) & 64 &2, 3, 5, 6 (1.33) &16, 17, 18, 13-2, 13-3 (0.15) & Guyomar and Lallart~\cite{Guy2011} & D& 1: review of nonlinear electronic interfaces for energy harvesting from mechanical vibrations, 2: comparative analysis of various switching techniques in terms of efficiency, performance under several excitation conditions, complexity of implementation \\
\hline\hline
\end{tabular}
\label{power}
}
\end{table*}

\subsection{MEMS/NEMS-based devices}
A large number of reviews on piezo-harvesters have been devoted to the field of MEMS/NEMS piezoelectric harvesters. Micro and nanoscale energy harvesters maybe useful at future for easy powering or charging of mobile electronics, even in remote areas, without the need for large power storage elements. MEMS-type devices include cantilever, cymbal and stack whereas NEMS type devices are wires, rods, fibers, belts, and tubes. Generation of output electric current using piezoelectric energy harvesters faces with many limitations and difficulties. Some of these limitations are low output power, high electric impedance, crack propagation in most piezoelectric materials due to overloading, frequency matching of the harvester with vibrational energy sources, and fabrication/integration of piezoelectrics in micro/nanoscale~\cite{selvan2016}.

\begin{figure}[htbp] \setlength{\unitlength}{1mm}
\hspace{0 in}
\begin{picture}(80,55)
\includegraphics[height=55mm]{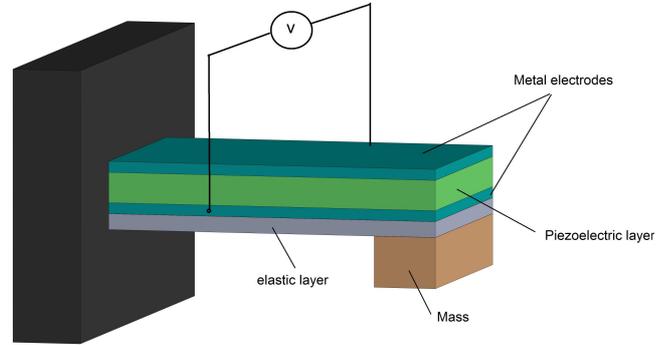}
\end{picture}
\caption {Unimorph structure of piezoelectric energy harvester that has one piezo-layer and a proof mass~\cite{kim2012}.}
\label{Fig:D}
\end{figure}

Kim et al.~\cite{kim2012} commented that for the elimination of chemical batteries and complex wiring in microsystems, a fully assembled energy harvester with the size of a US quarter dollar coin should be able to generate about 100$\mu$W of continuous power from ambient vibrations. In addition, the cost of the device should be sufficiently low. The article have addressed two important questions that are "how can one achieve self-powering when the power required is much larger than what can be achieved by MEMS-scale piezoelectric harvesters?” and "what is the best mechanism for converting mechanical energy into electrical energy at ~mm 3 dimensions?”. Also, they commented that for harvesting the power robustly, the resonance bandwidth of piezoelectric cantilevers should be wide enough to accommodate the uncertain variance of ambient vibrations. Thus, the resonance bandwidth is a significant characteristic for trapping an enough amount of energy onto the harvester and should be accounted for in determining the performance of energy harvesters.  MEMS technology is a cost-effective fabrication technology for PHEs if it can meet the requirements for power density and bandwidth. Three major aspects to make the MEMS PEHs appropriate for use in real applications are the final cost of the PEH, the normalized power density, and the operational frequency range (including the bandwidth and center frequency). They added that piezoelectric MEMS energy harvesters mostly have a unimorph cantilever configuration (Fig.~\ref{Fig:D}).  The proof mass (M) in Fig.~\ref{Fig:D} is used to adjust the resonant frequency to the available environmental frequency, normally below 100 Hz. Recently, integrated MEMS energy harvesters have been developed and in comparison of MEMS PEHs some essential merits like the active area of PEH, active volume, resonant frequency, harvested power, and power densities in volume or area, should be considered. They reviewed challenges of the piezo-harvesters, including the need to high power density and wide bandwidth of operation of the piezoelectric systems, the non-linear resonating beams for wide bandwidth resonance, and improvements in materials and the structural design. They concluded that the epitaxial growth and grain texturing of the piezo-materials, the embedded medical systems, the lead-free piezoelectric MEMS-based materials, and materials with giant piezoelectric coefficient  are active research fields. They presented an extensive comparison of thin-film piezo-systems from various sources and concluded that the state-of-the-art of power density is still about one order smaller than what is needed for practical applications.

Toprak and Tigli~\cite{top2014} conducted a review on piezoelectric harvesters based on their size (nanoscale, microscale, mesoscale, macroscale). They also presented an interesting statistics that the number of publications between 2009 and 2014 on piezoelectric harvesting is more than twice the sum of publications about the electromagnetic and electrostatic systems. They commented that the inherent reciprocal conversion capability is an important advantage of the piezoelectric energy harvesters that allows them to have simpler architectures in comparison to the electromagnetic and electrostatic counterparts. It is declared that the bio-compatibility, the reconciliation with the CMOS technology, the rectification and storage losses, and enhancing the operation bandwidth are the most challenging issues about such systems. A discussion on validity of the classical constitutive relations for the piezo-materials in nanoscale and pay attention to the minimum required power output of PEHs are felt missing in the paper.

Todaro et al.~\cite{todaro2017} reviewed the current status of the MEMS-based energy harvesters using piezoelectric thin films, and highlighted approaches and strategies. They commented that such harvesters are compact and cost-effective especially for harvesting energy from environmental vibrations. They believe that two main challenges of this topic to achieve high-performance devices are increasing the amount of generated power and the frequency bandwidth. They also introduced the theoretical principles and the main figures of merit of energy conversion in piezoelectric thin films. They compared most important thin film piezo-materials based on the introduced figure of merit. Their recommendations for future research are developing proper materials, new device architectures and strategies involving bimorph and multimorph designs exploited for bandwidth and power density improvements, progressing in synthesis and growth technologies for lead-free high quality piezoelectrics, employing new flexible materials with tailored mechanical properties for larger displacement and lower frequencies, and taking advantage of non-linear effects to obtain a wider bandwidth and a higher efficiency. Specifying the minimum required output power and attending to mechanical and electrical energy conversion efficiencies are felt missing in this review paper.

Dutoit et al. focused on design considerations for piezoelectric-based energy harvesters for MEMS-scale sensors. They stated that the power consumption of tens to hundreds of $\mu$W is predicted for sensor nodes and nowadays milli-scale commercial node has an average power consumption of 6–300 $\mu$W ~\cite{Dut2005}. With the reduction of power requirements for sensor nodes, the application of piezoelectric energy harvesters has become viable.They stated that the power or energy sources can be divided into two groups: sources with a fixed energy density (e.g. batteries) and sources with a fixed power density (normally ambient energy harvesters).
They suggested that the following information be made available in research papers to facilitate a relative comparison of PEH devices: device size, the maximum tip displacement at maximum power output, the mechanical damping ratio, the electrical load, the device mass, and the input vibration characteristics. Also, in this paper a fully coupled electromechanical model was developed to analyze the response of a piezoelectric energy harvester and the difference in optimization strategies od PEHs in on-resonant and off-resonant conditions were remarked.

Other review papers on MEMS PEHs have focused on several issues including ZnO nonorods and flexible substrates, and ZnO-base nano-devices  ~\cite{briscoe2015}, comparison of existing piezoelectric micro generators (including the impact coupled, the resonant and human-powered devices, and the cantilever-based setup) with electromagnetic and electrostatic mechanisms ~\cite{beby2006}, the description of micro and nano device fabrication techniques, performance metrics, and device characterization ~\cite{Gos2019}, hybrid electromagnetic-piezoelectric and triboelectric/piezoelectric MEMS-based harvesters and their privileges ~ \cite{Salim2015}, ZnO nanostructure-based photovoltaic, piezoelectric nano-generators, and the hybrid approach harvesting energy harvesting ~\cite{kumar2012}, reporting the benefits, capacities, applications, challenges, and constraints of micro-power harvesting methods using thermoelectric, thermophotovoltaic, piezoelectric, and microbial fuel cell ~\cite{selvan2016}, nanostructured polymer-based piezoelectric and triboelectric materials as flexible, lightweight, easy/cheap to fabricate, being lead-free,  biocompatible, and robust harvesters ~\cite{Jing2018},  theoretical and experimental characterization methods for predicting and determining the potential output of nano wire-based nanogenerators ~\cite{Wang2012}, reviewing the research progress in the field of piezoelectric nanogenerators and describing their working mechanism, modeling, and structural design ~\cite{wang2015}, discussing the impact of composition, orientation, and microstructures on piezoelectric properties of perovskite thin films like PbZr1-xTixO3 (PZT) in applications such as low-voltage radio frequency MEMS switches and resonators, actuators for millimeter-scale robotics, droplet ejectors, energy harvesters for unattended sensors, and medical imaging transducers ~\cite{Muralt2009}.

Table~\ref{nano} presents details of evaluation of reviews on micro/nanoscale energy harvesting. In summery, almost all review articles discussed some great challenges of development of MEMS/NEMS-based piezoelectric harvesters such as the limited bandwidth and low output power. On the other hand, there are some competitive technologies like electromagnetic, thermoelectric, and electrostatic energy harvesting that can be employed for scavenging the environment waste energy. Most of the comparative review papers have focused on the output power and coupling coefficient of the harvesting systems and other important features such as the lifetime, capability of working in harsh environmental conditions, the cost level, commercial accessibility, and the technology readiness level (TRL) need more deep considerations.

\begin{table*}
\caption{ Overall evaluation of review papers written on MEMS piezoelectric energy harvesters. The numbers in brackets denote the number of non-general future lines. "Cons." stands for conclusions. \\
\textbf{Conclusions:} Efficiency/performance improvement, necessity of frequency bandwidth broadening, necessity of optimizations, increasing life time and endurance, size reduction and manufacturability, importance of electric interface circuits, the necessity of material properties improvement. \\
\textbf{Merits:} 1: reporting the output power or power density of MEMS/NEMS PHEs, 2: coupling factors and operational mode, 3: describing the fabrication techniques, 4: matching the resonance frequency of PHEs with motivating frequencies, 5:paying attention to the minimum required output power,6: CMOS compatibility, 7: energy flow analysis, \\
\textbf{Sub-categories:} 1- Micro/ nano scale materials (1-1- Grain textured and epitaxial piezoelectric films, 1-2- Lead-free piezoelectric films, 1-3- Aluminum nitride piezoelectric film, 1-4- piezoelectric nano-polymers, 1-5- Polymer-Ceramic nanocomposite nano generators (NG), 1-6- Electrospun P(VDF-TrFE) nanofiber hybrid NGs, 1-7- Nylon nanowire-based piezoelectric NG, 1-8- Template-grown poly-L-lactic acid, 1-9- Electrospun poly-L-lactic acid nanofibers, 1-10- ZnO-polymer nanocomposite piezoelectric NG, 1-11- ZnO nano-rods, 1-12- Nano wires, 1-13- Nanowire-Composites, 1-14- PZT thin films, 1-15- piezo-polymer thin films, 1-16- piezoelectric electroactive polymers), 2- Nonlinear resonance-based energy harvesting structures, 3- energy conversion efficiency, 4- figure of merit for MEMS PHEs, 5- Material synthesis and deposition (5-1- solution phase synthesis, 5-2- thin film deposition, 5-3- growth of polymer-based nanowires), 6- Modes of operations for MEMS PHEs, 7- design configurations for MEMS PHEs (7-1- cantilever based piezoelectric generators, 7-2- other types of piezoelectric generators), 8- Microscale scale PHEs, 9- Substrate and electrode and their impact on performance, 10- MEMS device performance parameters, 11- Characterization of MEMS PHEs, 12- MEMS hybrid harvesters (12-1 Architectures of hybrid harvesters, 12-2- Mathematical models of (PZT hybrid harvesters, 12-3 PZT - Tribo-electric hybrid harvester), 13- Nano-scale PHEs (13-1 working principles, 13-2- design, fabrication and implementation of nanogenerators, 13-3- Hybrid nano-generators, 13-4- nano-rod arrays, 13-5- flexible nano generators, 13-6- ZnO nano-PHEs, 13-7- applications of nano-generators, 13-8- Flexoelectric enhancement at the nanometer scale, 13-9- Characterization of piezoelectric potential from piezoelectric NWs, 13-10- prototypes of nano generators, 13-11- Prediction of the power output from piezoelectric NWs, 13-12- vertically aligned nanowire arrays and their fabrication, 13-13- laterally aligned nanowire arrays and their fabrication), 14- Impact coupled devices, 15- Human powered piezoelectric generation, 16- Evolving technology of miniature power harvesters, 17- Positive prospects of micro-scale electricity harvesters, 18- Challenges and constraints of minute-scale energy harvesters, 19- CMOS compatibility, 20- biocompatibility, 21- bandwidth of PHEs, 22- figure of merit for PHEs, 23- piezoelectric thin films, 24- screening effect in PHEs, 25- Energy harvesting by piezoelectric thin films.
}
\resizebox{1.05 \textwidth}{!}{
\begin{tabular}{c|p{2.0cm}|p{0.7cm}|p{2.0cm}|p{1.7cm}|p{2.6cm}|p{1.0cm}|p{15.5cm}} \hline\hline
\# Cons. & Minimum required output & \# Refs. & Merits & Sub-categories & Ref. & Grade & Highlights \\ \hline\hline
6(0.85)&~$\mu W$ (1)& 89 &1, 2, 3, 4, 5, 6, 7 (2) &1 to 4 (0.25) & Kim et al.~\cite{kim2012}& A & Describing figure of merits for MEMS PHEs, Mentioning the key attributes for MEMS PHEs, Describing minimum acceptable power density for MEMS PHEs\\
4(0.57)&~$\mu W$ (0)& 95 & 1, 3, 4 (0.6) &1-11, 13-4, 13-5, 13-6 (0.15) & Briscoe and Dunn~\cite{briscoe2015} & C & 1: This review has summarized the work to date on nanostructured piezoelectric energy harvesters. 2: They stated that in order to satisfy the needs of real power delivery, devices need to maximize the rate of change of any strain delivered into a system in order to increase the polarization developed by the functional layers, and improve the coupling of the device to the environment. \\
4(0.57) &~$\mu W$ to $mW (0)$&123 &1, 2, 3, 4, 5, 6 (1.71) & 8, 13, 18, 19, 20, 21 (0.25) & Toprak and Tigli~\cite{top2014} & C & 1: They commented that the size-based classification provides a reliable and effective basis to study various piezoelectric energy harvesters. 2: They discussed the most prominent challenges in piezoelectric energy harvesting and the studies focusing on these challenges.\\
4(0.57)&~$\mu W$ to $mW (0)$&145 & 1, 2, 3, 4, 5 , 6 (1.71) &8, 18, 22, 23 (0.15) &Todaro et al.~\cite{todaro2017} & C &1: The paper has reviewed the current status of MEMS energy harvesters based on piezoelectric thin films. 2: The paper has highlighted approaches/strategies to face the two main challenges to be addressed for high performance devices, namely generated power and frequency bandwidth. 3: Comparison of lots of MEMS energy harvesters performances has been performed.\\
5(0.71)&$mW$ (0)&34 & 1, 2, 3, 4, 5 (1.4) & 12 (12-1 to 12-3) (0.1) & Salim et al.~\cite{Salim2015}& C
&Elaborating on the hybrid energy harvesters, reported Literature on such harvesters for recent years with different architectures, models, and results comparison of the present hybrid PHEs in terms of output power.\\
6(0.85)&~$\mu W$ to $mW (0)$& 74 & 1, 2, 4, 5 (1.15) & 6, 7, 8, 10, 25 (0.2) & Dutoit et al.~\cite{Dut2005} & C & Commenting on the necessary information for comparing different PHEs. Pointing on the difference between dominant damping components at the micro- vs. macro-scale. Developing a fully coupled electromechanical model for analyzing the response of PHEs with cantilever configuration.\\
7(1) &~$\mu W$ to $mW (0)$& 108 & 1, 3, 4, 5 (1.15) & 1-4 to 1-10, 5-3 (0.1) & Jing and Kar-Narayan~\cite{Jing2018} & C & 1: Discussing the growth of nanomaterials including nanowires of polymers of polyvinylidene fluoride and its co-polymers, Nylon-11, and poly-lactic acid for scalable piezoelectric and triboelectric nanogenerator applications. 2: discussing design and performance of polymer-ceramic nanocomposite.\\
6(0.85)&~$\mu W$ to $mW (0)$& 115 &1, 2, 4, 5 (1.15) & 7-1, 7-2, 14, 15 (0.2) & Beeby~\cite{beby2006} & C & Characterization and comparison of piezoelectric, electromagnetic and electrostatic MEMS generators\\
5(0.71)&~$\mu W$ (0) &140 & 1, 3, 4, 5 (1.15) & 1-2, 1-15, 1-16 (0.15) & Asif Khan~\cite{Khan2016} & C & 1: The review has covered the available material forms and applications of piezoelectric thin films. 2: The electromechanical properties and performances of piezoelectric films have been compared and their suitability for particular applications were reported. 3: Control over the growth of the piezoelectric thin films and lead-free compositions of thin films can lead to good environmental stability and responses, coupled with higher piezoelectric coupling coefficients.\\
3(0.4)&- (0) &75 &1, 3, 4, 5 (1.15) &1-15, 25 (0.1) & Muralt et al.~\cite{Muralt2009} & D& The article has reviewed the impact of composition, orientation, and microstructure on the piezoelectric properties of perovskite thin films
The author described useful power levels for MEMS PHEs.\\
5(0.71)&~$\mu W$ (0)&78 & 1, 2, 3 (0.85) & 1-12, 1-13, 13-7, 13-12, 13-13, 18 (0.25) &Wang et al.~\cite{wang2015}& D &  The working mechanism, modeling, and structure design of piezoelectric nanogenerators were discussed. Integration of nanogenerators for high output power sources, the structural design for increasing the energy harvesting efficiency in different conditions, and the development of practicable integrated self-powered systems with improved stability and reliability are the critical issues in the field classification of nano generators based on their desing and working modes were performed.\\
6(0.85) &$mW$ (0)&112 &1, 4, 5 (0.85) & 3, 16, 17, 18 (0.15) &Selvan and Ali~\cite{selvan2016} & D & The capabilities and efficiencies off our micro-power harvesting methods including thermoelectric, thermo-photovoltaic ,piezoelectric, and microbial fuel cell renewable power generators are thoroughly reviewed and reported\\
4(0.57) & ~$\mu W$ (0)& 69 & 1, 2, 4 (0.85) & 1-12, 13-8 to 13-11, 18 (0.15) & Wang~\cite{Wang2012} & D & 1: theoretical calculations and experimental characterization methods for predicting or determining the piezoelectric potential output of NWs were reviewed. 2: numerical calculation of the energy output from NW-based NGs. 3: Integration of a large number of ZnO NWs was demonstrated as an effective pathway for improving the output power.\\
4(0.57)&-(0) & 80 & 2, 3 (0.57) & 5 to 11 (0.3) & Gosavi and Balpande~\cite{Gos2019} & D & Description of some of synthesis and deposition techniques and performance parameters for MEMS PHEs\\
5(0.71) &-(0) & 100 & 1, 3 (0.57) & 13 (13-1 to 13-3) (0.05) & Kumar and Kim~\cite{kumar2012}& D &1: Describing the mechanism of power generation behavior of nano-generators fabricated from ZnO nanostructures, 2: describing an innovative and important hybrid approach based on ZnO nano-structures.
\\
\hline\hline
\end{tabular}
}
\label{nano}
\end{table*}

\subsection{Modeling approaches}
Some review papers have focused on the modeling of PHEs to clarify the physical bases behind the piezoelectric energy harvesting. There are a few number of review papers that have totally focused on evaluation of different modeling approaches for piezoelectric energy harvesting.

Erturk and Inman investigated mechanical~\cite{erturk200808} and mathematical~\cite{erturk2008} aspects of the cantilevered piezoelectric energy harvesters to avoid reuse of simple and incorrect older models in literature. They reviewed the general solution of the base excitation problem for transverse and longitudinal vibrations of a cantilevered Euler-Bernoulli beam. They proved that the classical single-degree-of-freedom (SODF) predictions may yield highly inaccurate results, and they are just appropriate for high tip-mass-to-beam-mass ratios. Damping due to internal friction (the Kelvin-Voigt damping), damping related to the fluid medium, the base excitation as a forcing function, and the backward piezoelectric coupling in the beam equation are among modeling parameters. Modelling of energy conversion efficiency is felt missing in the article.

Zhao et al.~\cite{Zhao2013} compared different modeling approaches for harvesting the wind energy, including the single-degree-of-freedom, the single-mode and multi-mode Euler-Bernoulli distributed-parameter models (ignored in Ref.~\cite{erturk2008}). They concluded that the distributed-parameter model has a more rational representation of aerodynamic forces, while the SDOF model more precisely predicts the cut-in wind speed and the electro-aeroelastic behavior. In addition, they performed a parametric study on the effect of the load resistance, wind exposure area, mass of the bluff body, and the length of the piezoelectric sheet on the cut-in wind speed as well as the output power level of the GPEH. Again, modelling of energy conversion efficiency is felt missing in the article.

Wei and Jing~\cite{Wei2017} presented a state-of-the-art review of theory, modeling, and realization of the piezoelectric, electromagnetic, and electrostatic energy harvesters. The linear inertia-based theory and the non-linear models have been described for three mentioned vibration-to-electricity converters. They investigated some characteristics of the piezo-harvesters such as being unaffected from external/internal electromagnetic waves, simple structure, depolarization, brittleness of the bulk piezo-layer, the poor coupling in piezo-film, and the poor adhesion with the electrode materials. Development of new piezoelectric materials, creation of new energy harvesting configurations by exploring the non-linear benefits, and design of efficient energy harvesting interface circuits are among their suggestions as future prospects. They concluded that the non-linearity is an important and effective parameter in terms of performance enhancement. Theoretical modeling of the non-linear systems with keeping reliability and stability is a challenging task. The reviewed models have not been compared in the paper.



Table~\ref{modelling} sums up the results of evaluation of the review papers written about the modelling approaches. The table also contains different sub-categories, the range of output power, the number of reviewed articles, the merits, general conclusions, and some other extra descriptions. The rank of each paper has been computed based on the number of merits, the number of subcategories,  the number of concluding remarks, and clear emphasizing on Svalue of minimum required output power.

\begin{table*}
\caption{{\normalsize Overall evaluation of review papers written on modeling of piezoelectric energy harvesters. The numbers in brackets denote the number of non-general future lines. "Cons." stands for conclusions. \\
\textbf{Conclusions:} Efficiency/performance improvement, necessity of frequency bandwidth broadening, necessity of optimizations, increasing life time and endurance, size reduction and manufacturability, necessity to improve the accuracy of PHE models. \\
\textbf{Merits:} 1: giving the mathematical background of the models, 2: considering the energy losses, 3: taking into account the resonance and off-resonance conditions, 4: efficiency modeling, 5: mentioning the constraints and limitations of the models, 6: mentioning the assumptions followed by the models, 7: comparison of the existing models. \\
\textbf{Sub-categories:} 1- Energy conversion in PHEs with linear models, 2- Energy conversion in PHEs with nonlinear models, 3- Modelling efficiency, 4- Modelling cantilever PHEs (4-1- SDOF models 4-2- distributed parameter modeling), 5- modeling the aeroelastic energy harvesting, (5-1- Flutter in airfoil sections, 5-2- vortex-induced vibrations in circular cylinders, 5-3- Galloping in prismatic structures, 5-4- VIV-/cylinder-based aeroelastic energy harvesters, 5-5- Galloping-based aeroelastic energy harvesters,5-6- Wake galloping, 5-7- SDOF models 5-8- Euler-Bernoulli distributed parameter model).
}
}
\resizebox{1.05 \textwidth}{!}{
\begin{tabular}{c|p{2.5cm}|p{1.2cm}|p{2.5cm}|p{2.0cm}|p{2.2cm}|p{1.2cm}|p{6.5cm}} \hline\hline
\# Cons. & Minimum required output & \# Refs. & Merits & Sub-categories & Ref. & Grade & Highlights \\ \hline\hline
6(1) &$mW$ (0)& 21 &1, 2, 3, 4, 5, 6, 7 (2) &4-1 to 4-3 (0.2) &Erturk and Inman~\cite{erturk200808,erturk2008}& B &Issues of the correct formulation for piezoelectric coupling, correct physical modeling, use of low fidelity models, incorrect base motion \\
6(1) & $mW$ (0)& 48 & 1, 3, 4, 5, 6, 7 (1.7) & 5-7 to 5-8 (0.2) &Zhao et al.~\cite{Zhao2013} & C & Comparing the performance of the modeling methods for GPEH, including the SDOF model, and single mode and multimode Euler-Bernoulli distributed parameter models. \\
5(0.85) & ~$\mu W$ (0) & 204 & 1, 2, 4 (0.85) & 1, 2, 3, 4 (0.8) & Wei and Jing~\cite{Wei2017}& C &1 reviewing the energy conversion efficiency of some of the conversion mechanisms, 2 describing several configuration design for PHEs like cantilever structures, and uniform membrane structures. \\
6(1) &$mW$ (0) & 201 & 3-5 (0.6) & 5-1- to 5-6 (0.2) & Abdelkefi~\cite{abdel2016}& D &Qualitative and quantitative comparisons between existing flow-induced vibrations energy harvesters, describing some of the limitations of existing models and recommending some improvement for future\\
\hline\hline
\end{tabular}
\label{modelling}
}
\end{table*}

\section{Applications}
\subsection{Vibration}
Vibration is the most common source of energy for piezoelectric harvesters, since there is no need to convert the input energy to the mechanical energy to produce electricity in piezo-materials. Also, its abundance, accessibility, and ubiquity in environment, in addition to multiple possible transduction types have made it more attractive for energy harvesting applications. The response of piezoelectric materials to the employed vibrations depends on their electromechanical properties like the natural frequency, their geometry, the electromechanical coefficients, and the damping characteristics. The design strategies for such types of harvesters, performance enhancement methodologies, behavior of the energy harvesters in harsh environment, their fatigue life, and failure mode, and the conditioning electric circuits are some of the important issues that should be addressed in review papers.

\begin{figure}[htbp] \setlength{\unitlength}{1mm}
\hspace{-0.2in}
\begin{picture}(80,100)
\includegraphics[height=100mm]{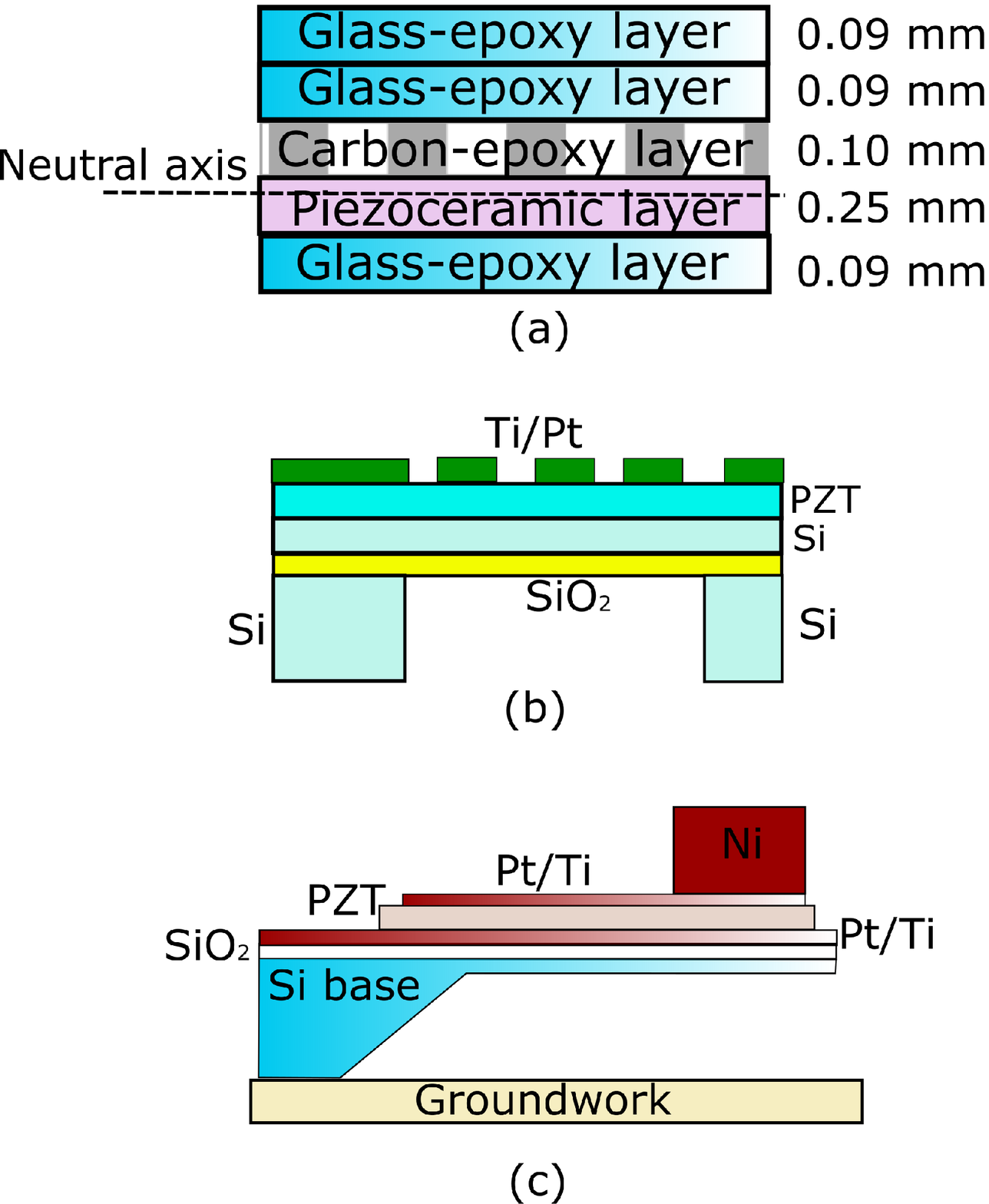}
\end{picture}
\caption {(a) Geometry and position of neutral axis of piezocomposite composed of layers
of carbon/epoxy, PZT ceramic and glass/epoxy~\cite{kim2011}, (b) a MEMS-based piezo-generator in 3-3 mode~\cite{sodano2004}, (c) schematic diagram of cross sectional view of a fabricated vibration-based micro power generator~\cite{Sid2015}.}
\label{Fig:vib}
\end{figure}

Kim et al.~\cite{kim2011} summarized the key ideas behind the performance evaluation of the piezoelectric energy harvesters based on vibration, classifications, materials, and the mathematical modeling of vibrational energy harvesting devices. They listed 17 important electro-mechanical characteristics of PZT-5H, PZT-8, PVDF, and described various configurations such as the cantilever type, the cymbal type, the stack type, and the shell type. They advised that the future opportunities for research are development of high coupling coefficient of piezoelectric materials, giving the ability to sustain under harsh vibrations and shocks, development of flexible and resilient piezoelectric materials, and designing efficient electronic circuitry for energy harvesters.
Siddique et al.~\cite{Sid2015} provided a literature review on vibration-based micropower generation using electromagnetic and piezoelectric transduction systems and hybrid configurations. They reported some performance characteristics of the piezoelectric energy harvesters with different materials and configurations. They claimed that most of the recent research have been devoted to modifications of the generator size, shape, and to introduce a power conditioning circuit to widen the frequency bandwidth of the system. Further research topics are development of the MEMS-based energy harvesters from renewable resources and making the miniature electric devices more reliable. Figure~\ref{Fig:vib} presents three schematic views of micorscale piezo-generators designed for vibration-based energy harvesting applications.

Sodano et al.~\cite{sodano2004}, as one of the earliest reviewers of the field, discussed the future goals that must be achieved for power harvesting systems to find their way towards the everyday use, and to generate sufficient energy to power the necessary electronic devices. They mentioned that the major limitations in the field of power harvesting revolve around the fact that the power generated by the piezoelectric energy harvesters is far too small to power most electronic devices. Increasing the amount of energy generation, developing innovative methods of accumulating the energy, use of rechargeable batteries, optimization of the power flow from a piezoelectric setup, minimizing the circuit losses, identifying the location of power harvesting and the excitation range, proper tuning of the power harvesting device are their predictions for future prospects of the vibration-based piezo harvesters.

Saadon and Sidek~\cite{saadon2011} presented a brief discussion of vibration-based MEMS piezoelectric energy harvesters. They summarized various designs of harvesters and reviewed  experimental results presented in the last 3 years before the date of publication of the paper. They focused on the working modes and maximum output power of the MEMS piezoelectric energy harvesters.
Harb~\cite{Harb2011} reviewed a brief history of all energy harvesting methods including the vibration-based, the electromagnetic-based, the thermal or radioactive-based, pressure gradient-based, the solar and light-based, biological, and micro-water flow systems.
However, it is advised that the different types of vibrations are the most available and the highest power provider sources.
The review papers like the one presented by Zhu et al.~\cite{Zhu2009} are the result of an explosive utilization of the vibration-based micro-generators in powering the wireless sensor networks. They demonstrated an overall review of the principles and the operating strategies to increase the operational frequency range of the vibration-based micro-generators.
Harne and Wang~\cite{Harne2013} reported the major efforts and findings about common analytical frameworks and principal results for bi-stable electromechanical dynamics, and a wide variety of bi-stable energy harvesters.
Based on their discussion, the remaining challenges of such systems are maintaining high-energy orbits, operation under stochastic vibratory conditions, designing the coupled bi-stable harvesters, and defining proper performance metrics.


In summery, different configurations of the piezoelectric cantilevers, their power output and the performance enhancement strategies have been covered by the review papers well. However, a systematic comparison of different configurations of piezoelectric energy harvesters, and also their ability to sustain harsh vibrations and shocks, their fatigue life, their cost and accessibility have not been considered by the reviews. Table~\ref{vibvib} presents the results of evaluation of the piezo-electric energy harvesters from vibrational sources. The table also contains different sub-categories, the range of output power, the number of reviewed articles, the merits, general conclusions, and some other extra descriptions. The rank of each paper has been computed based on the number of merits, the number of subcategories, the number of concluding remarks, and clear emphasizing on the value of minimum required output power.

\begin{table*}
\caption{{\normalsize Overall evaluation of review papers written on piezoelectric energy harvesting from vibration sources. "Cons." stands for conclusions.\\
\textbf{Conclusions:} 1: Efficiency/performance improvement, 2: frequency tuning, 3: safety issues, 4: costs, 5: hybrid harvesters, 6: non-linear models, 7: battery replacement, 8: miniaturization, 9: steady operation, 10: more efficient materials, 11: stochastic modeling. \\
\textbf{Merits:} 1: electromechanical coupling factor, 2: realistic resonance, 3: energy flow, 4: range of output. \\
\textbf{Sub-categories:} 1: circuits, 2: type of materials, 3: modeling, 4: noise level, 5: wearable, 6: frequency range, 7: MEMS}
}
\resizebox{1.0 \textwidth}{!}{
\begin{tabular}{c|p{2.5cm}|p{1.2cm}|p{2.5cm}|p{2.0cm}|p{2.2cm}|p{1.2cm}|p{6.5cm}} \hline\hline
\# Cons. & Minimum required output & \# Refs. & Merits & Sub-categories & Ref. & Grade & Highlights \\ \hline\hline
3 (0.27)& ~$\mu W$ to $mW$ (1) & 93 & 1-4 (2.0) & 1, 2, 3, 5 (0.57) & Kim et al.~\cite{kim2011} & B &Comparison with electrostatic and electromagnetic energy conversions \\ \hline
6 (0.55) & ~$\mu W$ to $mW$ (1) & 145 & 1-4 (2.0) & 7 (0.14) & Siddique et al.~\cite{Sid2015} &B &Comparison with electromagnetic and electrostatic \\\hline
6 (0.55) &0.17$\mu W$ (1) & 35 & 2-4 (1.5) & 1, 3, 4, 5 (0.57) & Sodano et al.~\cite{sodano2004} &B &Insufficient output power \\ \hline
3 (0.27) &60$\mu W$(1)  & 23 & 2, 4 (1.0) & 1, 7 (0.29) & Saadon and Sidek~\cite{saadon2011} & C& Inadequate output power \\ \hline 
- (0.0) &2.46$m W$ (1)  & 56 & 3, 4 (10.) & 1, 6 (0.29)& Harb~\cite{Harb2011} & C& From thermal sources, RF sources, CMOS devices, power management sources \\\hline
7 (0.64) &- (0) & 50 & 2, 3 (1.0) & 6 (0.14) & Zhu et al.~\cite{Zhu2009} &D &Focused on frequency tuning \\\hline
5 (0.46) &- (0) & 84 & 1, 2 (1.0) & 2, 3 (0.29) & Harne and Wang~\cite{Harne2013} &D&Focused on bistable systems, stochastic vibrations \\ \hline\hline 
\end{tabular}
\label{vibvib}
}
\end{table*}

\subsection{Biological sources}
Biomechanical energy harvesting provides an important alternative to electrical energy for portable electronic devices. Hwang et al.~\cite{hwang2015} addressed the developments of flexible piezoelectric energy-harvesting devices by using high-quality perovskite thin film and innovative flexible fabrication processes.
In addition, the energy harvesting devices with thick and rigid substrates are unsuitable for responding to the movements of internal organs and muscles.
They commented that the electric power harvested from the bending motion of a flexible thin film is sufficient to stimulate heart muscles. Also, Easy bendability, higher conversion efficiency, enhanced sensing capability in nanoscale, self-energy generation and the real-time diagnosis/therapy capabilities are among advantages of such systems.
Ali et al.~\cite{Ali2019} discussed the possibilities of utilizing the piezo-based energy conversion from the source of muscle relaxation and contraction, the body movement, the blood circulation, the lung and cardiac motion in applications such as pacemakers, blood pressure sensors, cardiac sensors, pulse sensors, deep brain simulations, biomimetic artificial hair cells, active pressure sensors, and active strain sensors. The piezoelectric materials containing nanowires, nanorods, nanotubes, nanoparticles, thin films, the lead-based ceramics, the lead-free ceramics, the polymer-based materials, the textured polycrystalline materials, and the biological piezo-materials have been evaluated. They proposed several challenging problems such as the flexibility to fit into the shape of an organ, the proper management of power, selection of a media for the electrical connection, enhancing the biological safety, designing the interface between the body tissue and the implanted piezo-material, efficient encapsulation, further miniaturization, and conducting related experiments on small/large animal and human cases.

Surmenev et al.~\cite{sur2019} described novel techniques in fabrication of hybrid piezoelectric polymer-based materials for biomedical energy harvesting applications such as detection of motion rate of humans, degradation of organic pollutants, and sterilization of bacteria. They described the different methods that can be employed for the improvement of the piezoelectric response of polymeric materials and scaffolds. They also reviewed biomedical devices and sensors based on hybrid piezo-composites. Similar to most other reviews, increasing the performance is one of proposed future works. Others are alignment of nanofiller particles inside the piezopolymer matrix, developing common standards for consistently quantifying and evaluating the performance of various types of piezoelectric materials, and investigation of the structural parameters.

The internal charging of implantable medical devices (IMD) is another important biological application of piezoelectric energy harvesting. Extending the lifespan of IMDs and the size minimization have become main challenges for their development. For such devices, energy from the body movement, muscle contraction/relaxation, cardiac/lung motions, and the blood circulation is used for powering medical devices. Zheng et al.~\cite{Zheng2017} presented an overall review of the piezoelectric energy devices in comparison to the triboelectric harvesters with the source of body movement, muscle contraction/relaxation, cardiac/lung motions, and the blood circulation. They proposed that future opportunities are fabrication of intelligent, flexible, stretchable, and fully biodegradable self-powered medical systems for monitoring biological signals, in vivo and in vitro treatment of various diseases, optimization of the output performance, obtaining higher sensitivity, elasticity, durability and biocompatibility, biodegradable transient electronics, intelligent control of dynamic properties in vivo, improving the operating lifetimes, and the absorption efficiency.
Mhetre et al.~\cite{Mheter2011} gave a brief review of micro energy harvesting techniques and methods from the limb movement for drug delivery purposes, dental applications, and the body heat recovery using the piezoelectric transducers. They just announced that the main challenge is to enhance the energy output using proper electronic circuit designs. Much more research is required to harvest energy from other biological parameters such as the body temperature and respiration.
An average amount of energy used by the body is $1.07\times10^7 J$  per day. This amount of energy is equivalent to approximately 800AA (2500mAh) batteries with the total weight of about 20 kg. This considerable amounts of human energy opens the road of development of energy harvesting technologies for powering electronic devices~\cite{Riem2011}. Riemer and Shapiro~\cite{Riem2011} investigated the amount of electricity that can be generated from motion of various parts of the body such as heel strike, ankle, knee, hip, shoulder, elbow, arm, leg, the center of mass vertical motion, and the body heat emission, using the piezo-harvesters and electrical induction generators.
They claimed that such technologies are appropriate for the third world countries, which is to some extent doubtful referring to low performance and high cost of fabrication.

\begin{figure}[htbp] \setlength{\unitlength}{1mm}
\hspace{-0.2in}
\begin{picture}(80,130)
\includegraphics[height=130mm]{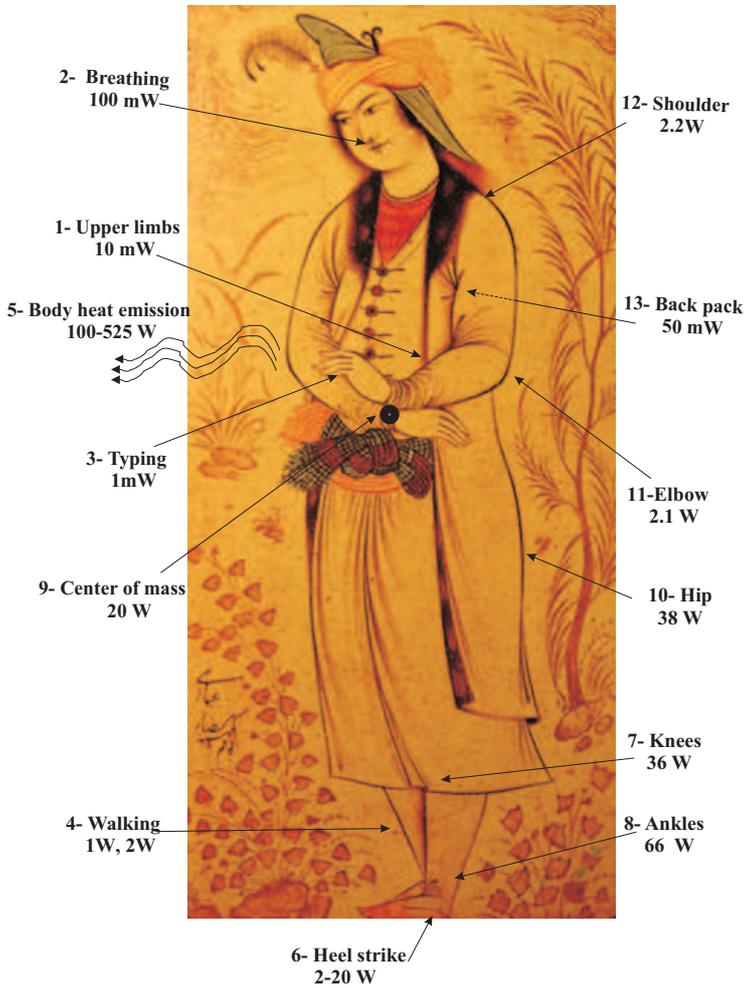}
\end{picture}
\caption {Available sources of energy from the human body organs. The data number 1-4 from Refs.~\cite{calio2014}, number 4-13 from~\cite{Riem2011}. The results are illustrated on Reza Abbasi's "Prince Muhammad-Beik" drawing (1620, public domain).}
\label{Fig:biodav}
\end{figure}

In addition to biocompatibility problems, the main challenges in development of these types of energy harvesters are constructing a device that can harvest as much energy as possible with minimal interference with the natural function of the body. Also, the device should not increase the amount of energy required by a person to perform his/her activities. Specially for IMDs, the lifetime and efficient power output of the energy harvesters are of outmost importance. Figure~\ref{Fig:biodav} illustrates magnitude of harvestable energy sources from the human body organs. Similar values can be predicted more or less from organs of animals in related applications.
Xin et al.~\cite{Xin2016} reviewed shoes-equipped piezoelectric energy harvesters. They described advantages and limitations of the current and newly developing piezoelectric materials, including the flat plate type, the arch type, the cantilever type, the nanocomposite-based, the photosensitive-based, and the hybrid piezoelectric-semiconductors technologies. They announced that enhancing the coupling coefficient of the piezoelectric materials and optimizing the structure of the energy harvester and the energy storing circuit require further investigation.

The reviewed articles about the biological applications have focused on highlighting new materials and structures of biological energy harvesters and their power output. The bio-compatibility, the interference of the device with the biological organ, reliability of the device along with its lifetime and economic issues are open topics in the field. Table~\ref{biobio} summarizes different highlights and descriptions of the review articles related to the biological applications. The grade of each paper has been computed based on the number of merits, the number of subcategories, the number of concluding remarks, and clear emphasizing on value of minimum required output power.

\begin{table*}
\caption{{\normalsize Overall evaluation of review papers written on piezoelectric energy harvesting from biological applications. "Cons." stands for conclusions.\\
\textbf{Conclusions:} 1: Efficiency/performance improvement, 2: safety issues, 3: costs, 4: hybrid harvesters, 5: non-linear models, 6: battery replacement, 7: miniaturization, 8: steady operation, 9:  efficient (flexible, stretchable, bio-compatible) materials, 10: self-powered, 11: being wearable, 12: control systems. \\
\textbf{Merits:} 1: electromechanical coupling factor, 2: realistic resonance, 3: energy flow, 4: range of output. \\
\textbf{Sub-categories:} 1: organ motion, 2: heel strike, 3: ankle, 4: Knee, 5: hip, 6: center of mass, 7: arms, 8: muscles, 9: cardiac/lung motion, 10: blood circulation, 11: heat emission, 12: drug delivery, 13: dental cases, 14: thin films, 15: artificial hair cell, 16: biosensors.}
}
\resizebox{1.0 \textwidth}{!}{
\begin{tabular}{c|p{2.5cm}|p{1.2cm}|p{2.5cm}|p{2.0cm}|p{2.2cm}|p{1.2cm}|p{6.5cm}} \hline\hline
\# Cons. & Minimum required output & \# Refs. & Merits & Sub-categories & Ref. & Grade & Highlights \\ \hline\hline
 5 (0.42)& 250 V, 8.7 $\mu$A (1)& 71 &1, 3, 4 (1.5)& 1, 9, 14, 15 (0.25) & Hwang et al.~\cite{hwang2015} & B&Focused on thin films \\\hline
 9 (0.75)& 11V, 283$\mu A$ (1)& 240 &1, 4 (1.0) & 1, 8, 9, 10, 16 (0.31) & Ali et al.~\cite{Ali2019} & B & - \\\hline 
 8 (0.67)& 1$\mu F$, 20$V$, 50s (1)& 235 &1, 4 (1.0) & 1, 16 (0.13) & Surmenev et al.~\cite{sur2019} & C& Lead-free polymer-based, size-dependent effects, insufficient output power of piezoelectric polymers and their copolymers \\\hline
 7 (0.58)& 11$mWcm^{-3}$ (1)& 107 &1, 4 (1.0) & 8, 9, 10 (0.19) & Zheng et al.~\cite{Zheng2017} & C &Comparison with triboelectric\\\hline
 1 (0.08)& ~mW(1)&29 &4 (0.5) & 12, 13 (0.13)& Mhetre et al.~\cite{Mheter2011} & C &-\\\hline 
 4 (0.33)& 2W (1)& 38 &4 (0.5) & 1, 2, 3, 4, 5, 6, 7, 11 (0.5) & Riemer and Shapiro~\cite{Riem2011} & C & Comparison with electrical induction generators and electroactive polymers \\\hline
 4 (0.33)& -(0)& 53 &1 (0.5) & 1 (0.06) & Xin et al.~\cite{Xin2016} & E & Shoes-equipped
Geometry classification \\\hline \hline
\end{tabular}
\label{biobio}
}
\end{table*}

\subsection{Fluids}
\begin{figure}[htbp] \setlength{\unitlength}{1mm}
\hspace{-0.0in}
\begin{picture}(80,50)
\includegraphics[height=50mm]{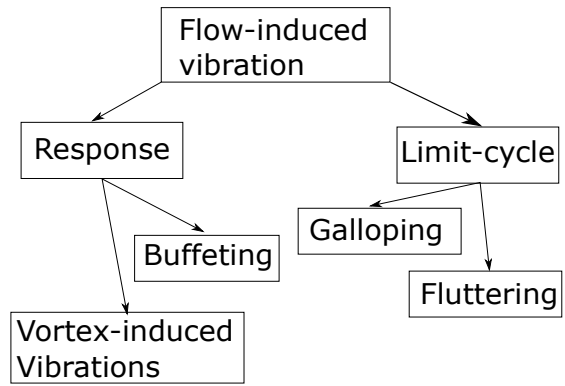}
\end{picture}
\caption {Different classes of energy harvesting categories from flow-induced vibrations~\cite{Wang2020}.}
\label{Fig:FIV}
\end{figure}

Wang et al.~\cite{Wang2020} categorized the fluid-induced vibrations for the purpose of energy harvesting into four categories based on different vibration mechanisms: the vortex induced vibration, galloping, fluttering, and buffeting. They discussed the vortex-induced vibrations and buffeting (as forced vibration cases), galloping and flutter (as limit-cycle vibration items) using electromagnetic, piezoelectric, electrostatic, dielectric, and triboelectric methods along with the corresponding numerical and experimental endeavors. They presented a fruitful summary of the current research status on flow-induced vibration hydro/aero energy harvesters. It is concluded that the flow pattern around bluff bodies, the size limitations, estimation of costs of equipment, the maintenance costs, the lifespan, protection of equipment in the case of extreme weather, possible environmental impacts, the non-linear modeling, the intelligent regulating elements such as artificial neural network, implementation of hybrid multi-purpose energy harvesters, and development of new materials need to be further studied.
Figure~\ref{Fig:FIV} presents four classes of energy harvesting: vortex-induced vibrations, buffeting, galloping, and fluttering, from vibration mechanisms corresponding to fluid flows~\cite{Wang2020}.

Truitt and Mahmoodi~\cite{Tru2013} reviewed effects of wind-based energy harvesting from flow-induced vibrations by bluff bodies and aeroelastic instabilities (fluttering and galloping). They presented an overall study of energy generation density and the peak power outputs versus the bandwidth. After a brief review of dynamics of piezoelectric energy harvesting, theories and principles, energy densities and output powers, they concluded that the balance of efficiency-cost-manufacturability is the future horizon of the topic. They suggested the use of PVDFs in fluid excitation applications due to their increased flexibility over PZTs. They concluded that the fluttering- and galloping-based methods generate a higher output power, but with a narrower frequency bandwidth in comparison to the vortex induced methods. Also, the final vision for energy harvesting may be active energy harvesting in which the system dynamics can actively change in real-time to meet changing environmental dynamics.
Viet et al.~\cite{viet2017} compared three energy harvesting methods, including electrostatic, electromagnetic, and piezoelectric technologies to indicate privileges of the piezoelectric harvesting in power generation, transmission, structural installation, and the economic costs. Then, they reviewed different design methodologies of harvesting energy from ocean waves.
Effects of longitudinal, bending, and shear couplings have been discussed. It is concluded that due to higher energy generation density, higher voltage generation capability, simpler configuration, and more economic benefits, the piezoelectric technology is superior to the other methods.
Elahi et al.~\cite{Elahi2018} studied the fluid-structure interaction-based, the human-based, and the vibration-based energy harvesting mechanisms by qualitatively and quantitatively analyzing the existing piezoelectric mechanisms. They reviewed the vortex-induced vibration, fluttering, galloping, and the human-related structures. They commented that a significant amount of research has been conducted on aeroelastic energy harvesters, but aerodynamic models can be improved by taking into account steady, quasi-steady, and unsteady aerodynamics.
McCarthy et al.~\cite{M2016} reviewed the research done on piezoelectric energy harvesting based on fluttering. They introduced the mathematical terms needed to define the performance of the fluttering harvester. They discussed effects of the Strouhal number as a function of the Reynolds number, the wind characteristics, and formation of the atmospheric boundary layer (ABL). They declared that the ultra-low power densities, the long return period of investments, and quantification and alleviation of the fatigue damage are the most challenges for fluttering energy harvesting. Based on their opinion, determining the fatigue life and some metrics for a piezoelectric flutter, weather and precipitation effects are active research fields.

Hamlehdar et al.~\cite{kasa2019} presented a review of energy harvesting from fluid flows. Despite the general topic of the paper, the piezo-energy harvesting from blood as a liquid has been ignored. They have performed a literature review on energy production from vortex induced vibration, the Karman vortex street, the flutter induced motion, galloping, and the waves with water and air as working fluids. Also, there is a short discussion on modeling challenges.
The results of the review conducted by Wong et al.~\cite{wong2015} implies that the piezoelectric energy harvesting form the rain drop has privileges such as simple structure, easier fabrication, reduced number of components, and direct conversion of vibrational energy to electrical charge.
They stated that the main challenge in this field is to design and optimize the raindrop harvester for outdoor uses, being resistant against sunlight, wind, the impact force of larger drops, being waterproof, showing appropriate sensitivity to drops, supplying constant-rate energy over long periods of time, and optimizing the power efficiency.
Chua et al.~\cite{Chua2016} reviewed different types of the raindrop kinetic energy piezoelectric harvester, including the bridge-structure, the cantilever structure with the impact point near the free-end, the cantilever structure with six impact points at varies surface locations, the cantilever structure with impact point at the center, the PVDF membrane or the PZT edge-anchored plate, and the collecting diaphragm cantilevers. Also, they presented a brief summary of characteristics of hybrid harvesters. It is stated that the best parameter to compare different harvesters is the efficiency rather than the output peak power. Then based on this criterion, it is found that the cantilever-type and the bridge-type energy harvesters made of PZT are the best choices. This is to some extent  in contrast to the recommendations of Wong et al.~\cite{wong2015}.

Table~\ref{fluidfluid} presents the details and highlights of review papers on fluid-based piezo-energy harvesting. The grade of each paper has been computed based on the number of merits, the number of subcategories, the number of concluding remarks, and clear emphasizing on value of minimum required output power.

\begin{table*}
\caption{{\normalsize Overall evaluation of review papers written on piezoelectric energy harvesting from fluids. The numbers in parentheses denote number of non-general future lines. "Cons." stands for conclusions.\\
\textbf{Conclusions:} 1: Efficiency/performance improvement, 2: frequency tuning, 3: safety issues, 4: costs, 5: hybrid harvesters, 6: non-linear models, 7: battery replacement, 8: miniaturization, 9: steady operation, 10: more efficient materials. \\
\textbf{Merits:} 1: electromechanical coupling factor, 2: realistic resonance, 3: energy flow, 4: range of output \\
\textbf{Sub-categories:} 1: water waves, 2: galloping, 3: fluttering, 4: buffeting, 5: modelling, 6: wind's vortex street, 7: instabilities, 8: raindrop, 9: mechanical design.}
}
\resizebox{1.0 \textwidth}{!}{
\begin{tabular}{c|p{2.5cm}|p{1.2cm}|p{2.5cm}|p{2.0cm}|p{2.2cm}|p{1.2cm}|p{6.5cm}} \hline\hline
\# Cons. & Minimum required output & \# Refs. & Merits & Sub-categories & Ref. & Grade & Highlights \\ \hline\hline
9 (0.9)& $>$0.0289mW (1) &125 & 1, 2, 3, 4 (2.0)& 1, 2, 3, 4 (0.44) & Wang et al.~\cite{Wang2020} &A& Internet of things, machine learning tools \\\hline 
4 (0.4)& 115mW(1) & 62 & 1\, 2, 3, 4 (2.0)& 2, 3, 6, 7 (0.44) & Truitt and Mahmoodi~\cite{Tru2013} & B& Active control theory \\\hline 
4 (0.4)& 116$\mu W/cm^3$ (1) & 96 & 1, 2, 4 (1.5) & 1, 9 (0.22)& Viet et al.~\cite{viet2017} &B & -\\\hline
5 (0.5)& ~nW to mW (1) & 256 & 1, 4 (1.0) & 2, 3, 6 (0.33) & Elahi et al.~\cite{Elahi2018} &C& Human-based sources \\\hline
5 (0.5)& 440$\mu W/cm^3$ (1) & 96 & 2, 4 (1.0) & 3, 6, 9 (0.33) & McCarthy et al.~\cite{M2016} &C & Noise level, Atmospheric boundary layer, Fatigue life \\\hline
5 (0.5)& $>$1$\mu$W(1) & 199 & 4 (0.5) & 1, 2, 5, 6, 7 (0.56) & Hamlehdar et al.~\cite{kasa2019} &C & Biomimetic design \\\hline
4 (0.4)& -(0) & 87 & 1, 3, 4 (1.5) & 8 (0.11) & Wong et al.~\cite{wong2015} &C & Size effects\\\hline
4 (0.4)& ~$\mu W$(1) & 73 & 4 (0.5)& 8 (0.11) & Chua et al.~\cite{Chua2016} & C& Circuit design, Hydrophilic surface \\\hline\hline
\end{tabular}
\label{fluidfluid}
}
\end{table*}

\subsection{Ambient waste energy sources}
Guo and Lu~\cite{gue2017} discussed recent advances in application of thermoelectric and piezoelectric energy harvesting technologies from the pavements. They found out that a pipe system cooperating with a thermoelectric generator is superior in terms of cost effectiveness and electricity output to piezoelectric transducers (fabricated with PZT). Based on their recommendations, the impact of the mentioned energy harvesting facilities to pavement performance, life cycle assessments, optimization with respect to traffic conditions and solar radiation, and the change in vehicle fuel consumption due to additional vehicle vibration or resistance should be evaluated in future works.
Duarte and Ferreira~\cite{duar2017} presented a comparative study of photovoltaic, thermoelectric, electromagnetic, hydraulic, pneumatic, electromechanical, and piezoelectric harvesting technologies. Evaluation parameters are the conversion efficiency, the maximum generated power, the installation method, and their TRL. They declared that the essential economic data of products are not yet available.
Wang et al.~\cite{wang2018} illustrated applications of the photovoltaic cells, solar collectors, geothermal, thermoelectric, electromagnetic, and piezoelectric energy extraction systems from bridges and roads in terms of energy output, benefit-cost ratio, and the technology readiness level.
Based on their conclusions, the grade of support of the piezoelectric harvesters by governments is low to medium, while the solar and geothermal systems are strongly being supported.
Pillai and Deenadayalan~\cite{pila2014} presented a review of acoustic energy harvesting methods and piezoelectricity as a promising technology in this category due to being sensitive and efficient at high frequency excitations.
they declared that optimization of the resonator and the coupling of thermo-acoustic engine to the acoustic-electricity conversion transducer are open research fields.
Khan and Izhar~\cite{Khan2015} reviewed the recent developments in the field of electromagnetic- and piezoelectric-based acoustic energy harvesting. They reported sound pressure levels of various ambient acoustic energy sources.
A set of useful data about the sound pressure level (dB) and the frequency of various acoustic energy sources have been reported. They declared that researchers were focusing on enhancing the performance of the piezoelectric membrane through novel fabrications and optimized geometrical configurations.
Duarte and Ferreira~\cite{Duart2016} made a comparative study on road pavement energy harvesting technologies.
They compared existing technologies based on the installed power (per area or volume), the conversion efficiency, the power density. Also, they classified the harvesting technologies based on their TRL (technology readiness levels) values. It is demonstrated that the piezoelectric technology is at high TRL grades. But, it delivers insufficient energy production rate with low economic characteristics.

Also, some of previously discussed papers have devoted a part of their review to piezo-based energy harvesting from waste energies. Performance of the electromagnetic- and piezoelectric-based vibration energy harvesters for energy production from bridges has been evaluated by Khan and Ahmad~\cite{KJhan2016}. 
They have expressed that the majority of current harvesters are constructed based on the electromagnetic effect, but the piezo-materials are commercially available and are easy to develop.
The resonant frequency is a critical parameter in such narrow-band low-frequency applications, which is a privilege of the electromagnetic systems.
Maghsoudi Nia et al.~\cite{nia2017} presented different technologies of converting the kinetic energy of the human body during walking to electricity by locating a harvesting system on the body or inserting a harvester in the floor.
In contrast to the results of Guo and Lu~\cite{gue2017}, it is recommended that the piezoelectric harvester is a better choice for such applications, due to simplicity and flexibility, regardless of a lower power output.
Yildirim et al.~\cite{yild2017} reviewed amplification techniques, resonance tuning methods, and non-linear oscillations in applications involving the ambient vibration harvesting, based on piezoelectric, electrostatic, and electromagnetic conversion methods.
Al-Yafeai et al.~\cite{Yad2020} reviewed methodologies to convert the dissipated energy in the suspension dampers of a car to electricity, along with discussing the mathematical car models and respective experimental setups. The disadvantages of the piezo-generator in comparison to other methods are poor coupling, high output impedance, charge leakage, and the low output current. However, the advantages are simple structure, no need to external voltage sources and mechanical constraints, compatibility with MEMS-based devices, high output power, and having wide frequency range.
Al-Yafeai et al.~\cite{Yad2020} presented a review of design considerations for energy harvesting from car suspension system, including different piezo-materials, various mathematical modeling, the power dissipation, the number of degree-of-freedoms, the road input, the location of the piezo-system, and the electronic circuit. Dagdeviren et al.~\cite{Dag2016} highlighted essential mechanical to electrical conversion processes and the key design considerations of flexible and stretchable piezoelectric energy harvesters appropriate for soft tissues of human body, smart robots and metrology tools. They declared that the development outlooks of such devices are the designs and fabrication techniques.

Table~\ref{wastew} presents details and highlights of the review papers on ambient and waste energy piezo-harvesting methods. The grade of each paper has been computed based on the number of merits, the number of subcategories, the number of concluding remarks, and clear emphasizing on value of minimum required output power.

\begin{table*}
\caption{{\normalsize Overall evaluation of review papers written on piezoelectric energy harvesting from waste energies. "Cons." stands for conclusions. \\
\textbf{Conclusions:} 1: Efficiency/performance improvement and optimization, 2: safety issues, 3: costs, 4: hybrid harvesters, 5: non-linear models, 6: battery replacement, 7: miniaturization, 8: steady operation, 9: efficient materials, 10: control systems. \\
\textbf{Merits:} 1: electromechanical coupling factor, 2: realistic resonance, 3: energy flow, 4: range of output \\
\textbf{Sub-categories:} 1: acoustic energy, 2: modelling, 3: road pavement, 4: railway, 5: bridge.}
}
\resizebox{1.0 \textwidth}{!}{
\begin{tabular}{c|p{2.5cm}|p{1.2cm}|p{2.5cm}|p{2.0cm}|p{2.2cm}|p{1.2cm}|p{6.5cm}} \hline\hline
\# Cons. & Minimum required output & \# Refs. & Merits & Sub-categories & Ref. & Grade & Highlights \\ \hline\hline
4 (0.4)& 241$Wh/y$ (1)&120&3, 4 (1.0)&3, 5 (0.4)&Wang et al.~\cite{wang2018}& C &Comparison with photovoltaic cell, solar collector, geothermal, thermoelectric, electromagnetic devices, Fatigue failure and life-cycle \\\hline
4 (0.4)& 100mW (1)& 65 & 3, 4 (1.0) & 2, 3 (0.4)& Guo and Lu~\cite{gue2017}& C & Comparison with thermoelectrics\\\hline 
4 (0.4)& 10-100W(1)&34&3, 4 (1.0)&4 (0.2)&Duarte and Ferreira~\cite{duar2017}& C & Comparison with electromagnetic devices, TRL level presented \\\hline
3 (0.3)&-(0)&80&2, 4 (1.0)&1, 2 (0.4)&Pillai and Deenadayalan~\cite{pila2014} & D &Comparison with thermo-acoustics \\\hline
2 (0.2)&-(0)&54&2, 4 (1.0)&1 (0.2)&Khan and Izhar~\cite{Khan2015}& D& Comparison with electromagnetics\\\hline
2 (0.2)&-(0)&97&4 (0.5)&3 (0.2)&Duarte and Ferreira~\cite{Duart2016}&E& Comparison with solar, thermoelectric, electromagnetic devices \\\hline\hline
\end{tabular}
\label{wastew}
}
\end{table*}

\section{Challenges and the roadmap for future research}
Table~\ref{stat3} illustrates the number of published review papers on each field, the year of the first and the last published review paper, and the research fund sources. It is obvious that the energy harvesting from ambient energies, the MEMS/NEMS and the fluid-based harvestinggs, and material considerations, respectively have the highest rate of publication of the review papers.  The numbers in brackets demonstrate the number of funded review papers. Although it is predictable that some supporters prefer to remain anonymous, it is seen that about 46\% of papers have been supported by a non-university organization. The last column of the table presents a list of organizations and respected countries that have devoted a full/partial financial support to the review papers on piezo-materials.

It is expected that the forthcoming review papers focus on specialized topics. However, they may still contain some degree of generality. Due to the multidisciplinary nature of the field, it is vital to publish comprehensive reviews on detailed aspects of the piezoelectric harvesters. Publication of review papers with general topics is not very welcomed anymore. The rate of publication of the review papers on biological topics is less than expected. Due to the rapid progress of piezoelectricity in biomedical engineering, increasing the number of reviews in related fields is inevitable. We suggest the researchers to present some state-of-the-art articles with specific topics including progress in piezoelectric materials, new applications of piezoelectric energy harvesters, and new developments in MEMS and NEMS piezoelectric harvesters.

  The results of comparative researches on energy harvesters for the railway demonstrated that, even in macroscale energy harvesting, the piezoelectric energy harvesters are not very successful with respect to other harvesting technologies. This situation may be worst for micro and nanoscale harvesters. We predict that the single (non-hybrid) piezoelectric energy harvesters would be the true choice only in some specific applications for which other harvesting systems have inherent limitations. Thus, there is an essential need for making fair comparisons of all types of energy harvesters for specific applications. On the other hand, we encounter the growing number of publications on piezoelectric energy harvesters. It should be noted that the real world selects the energy harvesting systems with higher performances and lower costs.

Based on the data listed in Table~\ref{stat3}, three types of research lines have been detected:
\begin{enumerate}
\item Pioneering topics that are still under consideration: general reviews (2005-2019), the design key points (2005-2020), the material-related studies (2009-2019), the MEMS-based devices (2006-219),
\item Pioneering topics without any recent publication of review papers: the modeling approaches (2008-2017), the vibration-based harvesters (2004-2015), sensors and actuators (2007-2016),
\item The newly developed topics: fluids (2013-2020), ambient waste energy (2014-2020), the biological applications (2011-2019).
\end{enumerate}

\begin{table*}
\begin{small}
\caption{\label{stat3} Statistics of review papers published on different topics related to piezoelectric energy harvesting. The numbers in brackets demonstrate the number of funded review papers in each field.}
\resizebox{1.0 \textwidth}{!}{
\begin{tabular}{cccccp{8.5cm}} \hline\hline
~\hfill \hfill~ &~\hfill Topic \hfill~&~\hfill \#reviews \hfill~& Period &\# reviews per year& Non-university research fund sources\\ \hline
~\hfill 1 \hfill~ &~\hfill General \hfill~&~\hfill 8(4) \hfill~&~\hfill 2005-2019 \hfill~&~\hfill 0.53\hfill~& National Science Foundation (USA), National Natural Science Foundation (China), Spanish Ministry of Science and Technology and the Regional
European Development Funds (European Union), NanoBioTouch European project/Telecom Italia/Scuola Superiore SantAnna (Italy). \\ \hline
~\hfill 2 \hfill~ &~\hfill Design \hfill~&~\hfill 15(5) \hfill~&~\hfill 2005-2020 \hfill~&~\hfill 0.94 \hfill~& Texas ARP (USA), U.S. Department of Energy Wind and Water Power Technologies Office (USA), Ministry of Higher Education (Malaysia), Natural Science and Engineering Research Council (Canada), National Natural Science Foundation (China)/EU Erasmus+ project/Bevilgning \\\hline
~\hfill 3 \hfill~ &~\hfill Material \hfill~&~\hfill 11(7) \hfill~&~\hfill 2009-2019 \hfill~&~\hfill 1.00 \hfill~& M/s Bharat Electronics Limited (India), National Nature Science Foundation (China), Office of Basic Energy Sciences, Department of Energy (USA)/Center for Integrated Smart Sensors funded by the Korea Ministry of Science (Korea), National Natural
Science Foundation (China)/Shanghai Municipal Education Commission and Shanghai Education Development Foundation (China), European Research Council/European Metrology Research Programme/ UK National Measurement System, National Natural Science
Foundation (China), China scholarship Council/China Ministry of Education/Institute of sound and vibration \\\hline
~\hfill 4 \hfill~ &~\hfill Modeling \hfill~&~\hfill 5(3) \hfill~&~\hfill 2008-2017 \hfill~&~\hfill 0.5 \hfill~& Air Force Office of Scientific Research (USA), Air Force Office of Scientific Research (USA), a NSFC project of China\\\hline
~\hfill 5 \hfill~ &~\hfill Vibration \hfill~&~\hfill 8(1) \hfill~&~\hfill 2004-2015 \hfill~&~\hfill 0.67 \hfill~& Energy Efficiency \& Resources of the Korea Institute of Energy Technology Evaluation/Creative Research Initiatives\\\hline
~\hfill 6 \hfill~ &~\hfill Biology \hfill~&~\hfill 6(5) \hfill~&~\hfill 2011-2019 \hfill~&~\hfill 0.67 \hfill~& Russian Science Foundation/Alexander von Humboldt Foundation/European Commission, National Key R\&D Project from Minister of Science and
Technology (China), Basic Science Research Program (Korea)/Center for Integrated Smart Sensors as Global Frontier Project, R\&D Center for Green Patrol Technologies through the R\&D for Global Top Environmental Technologies program funded by the Korean Ministry of Environment, Paul Ivanier Center for Robotics and Manufacturing Research/Pearlstone Center for Aeronautics Research\\ \hline
~\hfill 7 \hfill~ &~\hfill Sensors \hfill~&~\hfill 5(4) \hfill~&~\hfill 2007-2016 \hfill~&~\hfill 0.5 \hfill~& Spanish Ministry of Education and Science, NSSEFF/fellowship/ NSF/ Ben Franklin Technology PArtners/the Center for Dielectric Studies/ARO/DARPA/the Materials Research Institute/U.S Army Research Laboratory, Converging Research Center Program by the Ministry of Education Science and Technology (Korea), Basic Science Research Program through the National Research Foundation
of Korea\\\hline
~\hfill 8 \hfill~ &~\hfill MEMS/NEMS \hfill~&~\hfill 15(7) \hfill~&~\hfill 2006-2019 \hfill~&~\hfill 1.07 \hfill~& National Science Foundation (China), the Basic Science Research Program, through the National Research Foundation of Korea, European Research Council, Ministry of Education (Malaysia), Office of Basic Energy Sciences Department of Energy (USA), International Research and Development Program of the National Research Foundation of Korea\\\hline
~\hfill 9 \hfill~ &~\hfill Fluids \hfill~&~\hfill 8(3) \hfill~&~\hfill 2013-2020 \hfill~&~\hfill 1.00 \hfill~& Ministry of Higher Education (Malaysia), National Natural Science Foundation (China), Australian Research Council/FCSTPty Ltd\\\hline
~\hfill 10 \hfill~ &~\hfill Ambient \hfill~&~\hfill 11(4) \hfill~&~\hfill 2014-2020 \hfill~&~\hfill 1.57 \hfill~& Center for Advanced Infrastructure and Transportation (USA), Portuguese Foundation of Science and Technology, European Regional Development Fund, National Natural Science Foundation of China \\\hline
\end{tabular}
}
\end{small}
\end{table*}

The missing topics and the concluding future research topics, which need more close investigations to demonstrate their state-of-the-art are
\begin{enumerate}
\item Development of hybrid multi-purpose energy generators to completely harness energy of any kind and with any characteristics combining the piezo-pyro-tribo-flexo-thermo-photoelectric technologies.
\item Investigation of the mathematical models, the analytical and numerical solution techniques especially in nanoscale geometries where the classical continuum mechanics principal fails or in stochastic and non-linear situations. Some modified constitutive relations may need to be developed in non-continuum regimes. Also, the second law analysis and analysis of such systems from the thermodynamic viewpoint are the missing topics. The ab initio first principal simulations with atomistic nature are other challenging aspects of the nanoscale piezo-harvesters. Development of opensource codes like OpenFOAM and LAMMPS to include the solvers involving the piezoelectric effect may be another future research topic.
\item Application of piezo-materials in energy saving or reducing energy demand of a system rather than generation of energy requires a comprehensive review. An example of such energy reduction is the delay in decaying disturbances and delaying transition to turbulence using piezo-actuators placed on the surface of bluff bodies.
\item Due to the multi-physics nature of the piezoelectric effect, it is highly recommended to prepare review papers on optimization methods or machine learning-related topics.
\item Commercialization of the piezo-based harvesters and enhancing the technology readiness level need a serious attention. Perhaps, the next decade is the decade of extensive commercialization of the piezo-harvesters.
\item Plenty of patents have been published in recent years. Even some review papers should be devoted to investigation of patents presented in the field.
\item Focused reviews are needed on vibration-based piezo-harvesters in four recent years, development of piezotronics, and design of complete self-powered autonomous systems.
\item The overall design of devices including all parts, integrating the whole device in thin films, accumulation in rechargeable batteries, and taking into account the energy consumption needed to store the harvested energy.
\item Optimization of device architecture and size reducing configurations for portable applications, flexible wearable compact embedding implantable devices.
\item In situ prototype testing and design of harvesters coupled with environment and realistic applications to face with sunlight in outdoor applications, naturally occurring stochastic vibrations, the wind speed variation, dust, noises, required flexibility to fit the shape of human organs, and waterproofness.
\item Quantification of the figure of merit for the piezo-material properties such as energy transforming or conversion efficiency and standardizing the performance of piezo-based devices.
\item Reducing the maintenance cost, enhancing the lifespan, ameliorating the performance, analysis of government supports, the cost-benefit balance, and investigations of piezo-harvesting from energy policy viewpoint.
\item Thermal design of piezo-systems including the temperature-dependent properties and high-temperature harvesting limitations.
\item Fabrication of new piezo-materials with the non-linear behavior, larger displacements, lower frequencyies, wider operation bandwidth, and the frequency self-adaptation capabilities.
\item Use of meta-materials, non-toxic, biocompatible, printable piezo-materials, nanofibers, lead-free and high-piezoelectric coefficient materials.
\item Improving the design of electrical circuitry and managing rectification and storage losses.
\item Modification of structural designs including fracture-fatigue studies to increases reliability, stability, and durability of the device.
\item Design of efficient control techniques.
\item Extending the application of piezo-materials in novel fields such as internet of things.
\item Paying close attention to the use of unimorph design for high-energy harvesting rate, obtaining realistic resonance data in order to reach compactness, investigating energy output much lower than 1mW, and step-by-step report of successive energy flow or efficiency from input mechanical energy to the final electric energy in a rechargeable battery.
\item Focusing on applications involving elimination, restriction, and replacement of toxic materials and environmental pollution.
\item Development of designs exhibiting the highest electromechanical coupling factor.
\item Considering mechanical impedance matching, electromechanical transduction, electrical impedance matching and priority of these factors.
\item Development of other applications as energy harvesting devices with low energy demand.
\item Designing a grid of nano-devices (thousands) or thick films (10 to 30 microns) to generate minimum 1mW power (the required electric energy to operate a typical energy harvesting electric circuit with a DC/DC converter).
\item General development directions may be remote signal transmission, and energy saving in rechargeable batteries.
\end{enumerate}

However, the research on piezoelectric energy harvesting is not mature enough and many interdisciplinary active research fields are currently available. It should be mentioned that the progress of small-scale devices with very low power need is tightly tied to the revolution in design of efficient high-output power piezoelectric energy harvesters. It is recommended to lie on fundamental principles in order to obtain unique designs for future research.
\section*{Acknowledgment}
This research was supported by the Iran National Science Foundation (Grant number 98017606).

\end{document}